\documentclass[amsmath,amssymb,superscriptaddress, showpacs, showkeys,twocolumn, prx]{revtex4-1}

\usepackage{graphicx}
\usepackage{dcolumn}
\usepackage{bm}
\usepackage{amsmath}
\usepackage{amssymb,amsthm}
\usepackage{color}
\usepackage{epstopdf} 
\usepackage[sans]{dsfont}
\usepackage{accents}
\usepackage[utf8]{inputenc}
\usepackage[T1]{fontenc}
\usepackage[english]{babel}
\usepackage{hyperref}
\usepackage{mathbbol,bbm}
\usepackage{float}
\usepackage{accents}
\usepackage{mathtools} 

\newcommand\numberthis{\addtocounter{equation}{1}\tag{\theequation}}

\providecommand{\abs}[1]{\left|#1\right|}		
\newcommand{\tr}[1]{\ensuremath{\textrm{Tr}\left[#1\right]}}		
\newcommand{\Id}{\mathbb{1}}		
\newcommand{\imag}[1]{\ensuremath{\text{Im}\left[#1\right]}}		
\newcommand{\real}[1]{\ensuremath{\text{Re}\left[#1\right]}}			
\def\ket#1{\mathinner{|{#1}\rangle}}		
\newcommand{\braket}[2]{\langle #1|#2\rangle}		
\newcommand{\ketbra}[2]{\left|#1\rangle\langle #2\right|}		
\newcommand{\matelem}[3]{\langle #1|#2|#3\rangle}		
\newcommand{\op}[1]{\ensuremath{\hat{#1}}}
\newcommand{\avg}[1]{\ensuremath{\overline{#1}~}}
\newcommand{\ens}[1]{\ensuremath{\left\{#1\right\}}}
\newcommand{\commutator}[2]{\left[#1,#2\right]}
\newcommand{\subsum}[2]{\ensuremath{{\mathclap{\substack{#1\\ #2}}}}}
\newcommand{\limit}[2]{\underset{#1\rightarrow#2}{\lim}}

\newcommand{\eqdef}{:=}
\def\Re{\mathbb{R}}

\newcommand{\Eavg}{\ensuremath{\varepsilon}}
\newcommand{\Eno}{\ensuremath{E^0}}

\newcommand{\Pwidth}{\ensuremath{\sigma}}
\newcommand{\Ham}{\ensuremath{\op{\textrm{H}}}}

\newcommand{\Enslabel}{\ensuremath{\lambda}}
\newcommand{\Dm}{\ensuremath{\op{\rho}}}
\newcommand{\Dmvector}{\ensuremath{\vec{\rho}}}

\newcommand{\Avgdm}{\overline{\hat{\rho}}}

\newcommand{\Dmatrix}{\ensuremath{F}}

\newcommand{\Dim}{\ensuremath{d}}

\newcommand{\QMEmatrix}{\ensuremath{Q}}

\newcommand{\Lop}{\ensuremath{\op{L}}}

\newcommand{\Avgham}{\ensuremath{\avg{\Ham}}}

\newcommand{\Perturb}{\ensuremath{V}}
\newcommand{\Perturbop}{\ensuremath{\op{V}}}
\newcommand{\Perturbparam}{\ensuremath{\alpha}}
\newcommand{\Liouket}[1]{\ensuremath{\left|#1\right\}}}

\newcommand{\Liouketbra}[2]{\ensuremath{\left|#1\right\}\left\{#2\right|}}
\newcommand{\Lioumatelem}[3]{\ensuremath{\left\{#1\right|#2\left|#3\right\}}}
\newcommand{\Phase}{\ensuremath{\varphi}}
\newcommand{\Dmonly}{\ensuremath{\rho}}

\newcommand{\PrateA}{\ensuremath{\gamma}}
\newcommand{\PrateB}{\ensuremath{\Gamma}}
\newcommand{\PrateO}{\ensuremath{\Upsilon}}
\newcommand{\Charact}{\ensuremath{\phi^*}}
\newcommand{\Escale}{\omega_0}
\newcommand{\Projectionop}{\hat{\Pi}}

\begin{document}

\title{Protecting quantum coherences from static noise and disorder}

\author{Chahan M. Kropf} 
\email{Chahan.Kropf@gmail.com}
\affiliation{Istituto Nazionale di Fisica Nucleare, Sezione di Pavia, via Bassi 6, I-27100 Pavia, Italy
}
\affiliation{Department of Physics, Universit\`a Cattolica del Sacro Cuore, Brescia I-25121, Italy}
\affiliation{ILAMP (Interdisciplinary Laboratories for Advanced Materials Physics), Universit\`a Cattolica del Sacro Cuore, Brescia I-25121, Italy}

\keywords{Quantum Master Equations, Disorder, Static Noise, Quantum Coherences}

\date{\today}

\begin{abstract}
\noindent
Quantum coherences are paramount resources for applications, such as quantum-enhanced light-harvesting or quantum computing, which are fragile against environmental noise. We here derive generalized quantum master equations using perturbation theory in order to describe the effective ensemble-averaged time-evolution of finite-size quantum systems subject to static noise on all time scales. We then analyse the time-evolution of the coherences under energy broadening noise in a variety of systems characterized by both short and long-range interactions, by strongly correlated and fully uncorrelated noise  -- a single qubit, a lattice model with on-site disorder and a potential ladder, and bosons in a double-well potential with random interaction strength -- and show that couplings can partially protect the system from the ensemble-averaging induced loss of coherence. Our work suggests that suitably tuned couplings could be employed to counter-act part of the dephasing detrimental to quantum applications. Conversely, tailored noise distributions can be utilized to reach target non-equilibrium quantum states.
\end{abstract}

\maketitle


\section{Introduction}

For applications such as quantum-enhanced efficient light-harvesting \cite{Scholes2011} or quantum computing \cite{Nielsen2010} coherences are essential resources that in general are fragile against environmental noise \cite{Suter2016}. Among these, the static noise sources (often also termed disorder) are typically seen as technological imperfections, whose impact on the effective dynamics of the coherences can however be dramatic \cite{Kropf2016a}. The latter can also be actively exploited, for instance for random lasing \cite{Wiersma2008}, channel coding of a measurement \cite{Kechrimparis2019a}, or novel multi-component materials with unprecedented properties \cite{Toher2019}. Hence, a deep understanding of the effects of the static noise on the coherences, and in particular their time-evolution, is crucial for the development of quantum technologies \cite{Suter2016}. 

Characterizing the dynamics of noisy systems requires a statistical approach, as single realizations cannot give reliable predictions on reproducible features of the system as a whole. In the case of static noise, one studies an ensemble of unitary trajectories each one characterized by a random time-independent Hamiltonian. The most natural and accessible quantity is the average over this ensemble, since it is often difficult or even impossible to measure single realizations for several successive experimental runs as would be required to access higher-order correlations. As was established and extensively discussed in Ref.~\cite{Kropf2016a}, the ensemble-averaging procedure implies an average over the accumulated phases associated with the eigenstates of every realization of the underlying random Hamiltonian which induces a loss of phase information, hence dephasing, i.e., a specific type of decoherence \cite{Nielsen2010, Breuer2002}. In other words, the ensemble-averaged state must be described in terms of a density matrix as it evolves non-unitarily, in contrast to the state of individual realizations of the noise which evolve unitarily. A simple example of averaging-induced decoherence is the decay of the transverse-polarization in nuclear magnetic resonance experiments due to magnetic field inhomogeneities \cite{Slichter1990}. Note that static noise can only give rise to unital (non-unitary) ensemble-averaged dynamics, which excludes certain types of decoherence such as amplitude damping.

\begin{figure}
	\includegraphics[width = \linewidth]{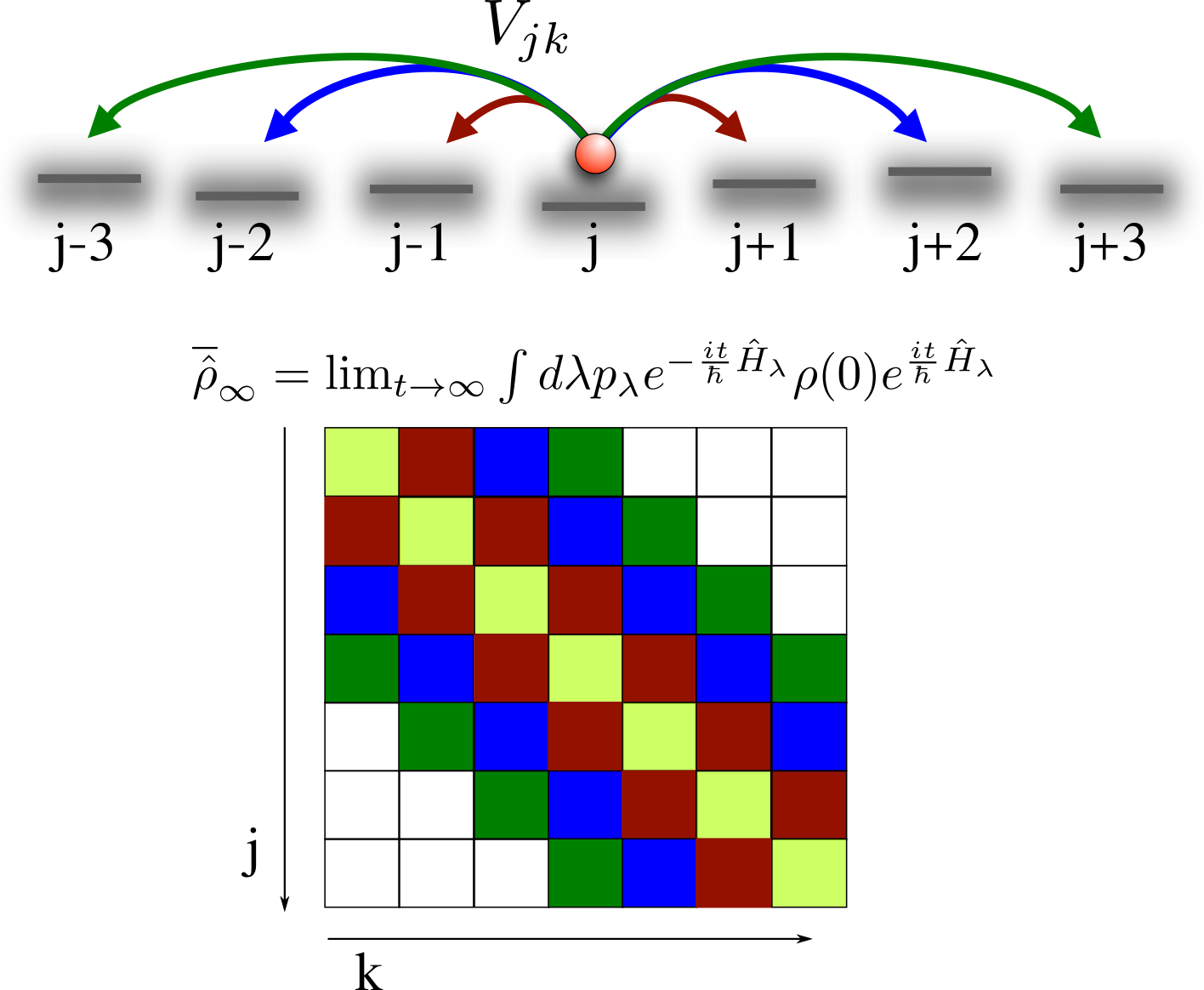}
	\caption{Illustration: Static noise in the on-site energies of a chain leads to a decay of coherences $\avg{\rho}_{jk}$ (off-diagonal elements) of the ensemble-averaged density matrix, but long-range coupling $\Perturbop$ can prevent a complete decay in the asymptotic state $\Avgdm_\infty$.}
	\label{fig:sketch}
\end{figure} 

An appropriate theoretical framework to describe the measurable ensemble-averaged dynamics, and in particular the evolution of the coherences (off-diagonal elements of the density matrix in a specific basis), is provided by generalized (quantum) master equations \cite{Kropf2016a}. We here present a master equation that describes the ensemble-average dynamics of finite disordered quantum systems that can be treated with time-independent perturbation theory in the absence of noise. Note that the small parameter needs not be specified more precisely ; it is sufficient to assume that the system in question is well described by standard perturbation theory. Hence, our approach is quite general and allows to describe the dynamics of a large variety of noisy systems in terms of effective, physically meaningful (time-dependent) shifts and (time-dependent) rates. A key point of the approach is that the perturbative expansion is done on the level of the state's representation -- more precisely, on the level of the dynamical matrix which contains the same information as the Choi matrix \cite{Choi1975} --, and \emph{not} on the level of the (Von Neumann) equation of motion for the density matrix. We thus not only obtain a description of the ensemble-averaged dynamics on transient times, but also of the asymptotic state.

We will derive perturbative master equations to study the time-evolution of the coherences in a variety of systems with static noise in the energies (diagonal noise) -- a broadened two-level system, a lattice model with on-site disorder, and bosons in a double-well potential with random interaction strength. We demonstrate that the coherences of the levels coupled by the perturbation are protected from complete averaging-induced decoherence (see illustration Fig.~\ref{fig:sketch}). As we will discuss in more details, this occurs because the ensemble-averaged dynamics describe a dephasing process in the eigenbasis of the noise-free Hamiltonian, and consequently the asymptotic state is the projection of the initial state onto the eigenbasis of this Hamiltonian, which may then exhibit non-vanishing coherences.

The article is organized as follows. The static noise model and the ensemble-averaged dynamics are defined in Section \ref{sec:System}. The general perturbative master equation is derived in Section \ref{sec:PME derivation}, and the particular form for diagonal static noise in Section \ref{sec:PME for on side disorder}. The effects of the noise on the dynamics of the coherences is studied for various systems in Section \ref{sec:LongRange}, and Section \ref{sec:Conclusions} concludes.

\section{Static noise and ensemble-averaged dynamics}\label{sec:System}

Our model considers quantum systems that can effectively be described by an ensemble of finite-size random $d$-dimensional Hamiltonians \cite{Kropf2017a}, which accounts for the descriptions of a large variety of experimental scenarios -- quantum systems coupled to a classical environment \cite{Chen2017, Breuer2018, Kropf2017a}, slow fluctuations in the experimental system parameters over time \cite{Sorelli2019, Kechrimparis2019} or generic statistical distributions of parameters usually referred to as disorder \cite{Anderson1958,Menon2012,Vojta2019}. Here we focus on the particular case that for each realisation of the noise, the system is described by a random Hamiltonian $\Ham_\Enslabel$ with an unperturbed, diagonalizable part $\Ham^{(0)}_\Enslabel$, and a perturbation $\Perturbparam\op{\Perturb}_\Enslabel$, where $\Perturbparam \ll 1$ is a dimensionless perturbation parameter and $\Enslabel$ labels the individual realisations. The random Hamiltonian ensemble is then characterized by
\begin{align}
	\ens{\Ham_\Enslabel = \Ham^{(0)}_\Enslabel + \Perturbparam \op{\Perturb}_\Enslabel \;\; , \;\; p_\Enslabel} \label{eq:Perturb general ensemble}
\end{align}
with $p_\Enslabel$ the probability density associated with the realization $\Enslabel$. The ensemble-averaged dynamics are characterized by the average density matrix
\begin{align}\label{eq:EnsAvgDyn}
	\Avgdm(t) =\int d\Enslabel p_\Enslabel e^{-\frac{it}{\hbar} \Ham_\Enslabel} \rho(0) e^{\frac{it}{\hbar} \Ham_\Enslabel},
\end{align}
where $\Dm(0)$ is the initial state (independent of the realisation of the noise), and overline $\overline{\cdot}$ denotes ensemble-averaged quantities. 
Note that in this manuscript we will use the terms static noise and disorder interchangeably, as in our context they equally give rise to a statistical ensemble description of the form Eq.~\eqref{eq:Perturb general ensemble}. 
 
For simplicity, we further assume that the unperturbed Hamiltonians $\Ham_\Enslabel^{0}$ are diagonal in the same eigenbasis $\ens{\ket{j}}_{j=1}^\Dim$ for all realisations of the noise. This assumption is for convenience, and serves to identify a specific basis in which to represent the master equation. We thus consider Hamiltonians consisting of a diagonal term $\Ham^{0}_\Enslabel =\sum_{j=1}^\Dim \Eno_{\Enslabel,j} \ketbra{j}{j}$ and a small coupling potential $ \Perturbparam \op{\Perturb}_\Enslabel = \sum_{j,k=1}^\Dim \Perturb_{jk}^\Enslabel \ketbra{j}{k}$.


\section{Derivation of the perturbative master equation}\label{sec:PME derivation}
In this section we derive a perturbative master equation characterized the ensemble-averaged dynamics, Eq.~\eqref{eq:EnsAvgDyn}, of the random ensemble, Eq.~\eqref{eq:Perturb general ensemble}. As a starting point we derive the (non-degenerate) perturbative expansion of the unitary dynamics \cite{LeBellac2006,Sakurai2011} for a single realisation of the static noise and build the associated perturbative dynamical matrix \cite{Sudarshan1961,Kropf2016a}.  

The unitary dynamics in Liouville space $\Liouket{\rho_\Enslabel(t)} = \mathcal{U}_\Enslabel(t) \Liouket{\rho(0)}$  associated with one realisation of the Hamiltonian $\Ham_\Enslabel = \Ham^0_\Enslabel + \Perturbparam \Perturbop_\Enslabel$ are characterized by the time-evolution superoperator
\begin{align}
	\mathcal{U}_\Enslabel(t) \Liouket{\rho(0)} = \op{U}_\Enslabel(t) \Dm(0) \op{U}_\Enslabel^\dagger(t) =  e^{-\frac{it}{\hbar} \Ham_\Enslabel} \Dm(0) e^{+\frac{it}{\hbar} \Ham_\Enslabel} 
\end{align}

For each realization of the random Hamiltonian $\Ham_\Enslabel$,  the Livouille time-evolution superoperator $\mathcal{U}_\Enslabel(t)$ is expanded in a time-ordered series \cite{Kropf2017}
 \begin{align}
 	  &\mathcal{U}_\Enslabel(t) =\mathcal{U}^0_\Enslabel(t) -\left(\frac{i\Perturbparam}{\hbar}\right) \int_0^t \, dt' \, \mathcal{U}^0_\Enslabel(t-t') \mathcal{\Perturb}_\Enslabel \mathcal{U}^0_\Enslabel(t') \label{eq:Perturbation expansion Liouville}\\
 	 &+\left(-i\frac{\Perturbparam}{\hbar}\right)^2 \int_0^{t} dt' \int_0^{t'} dt'' \, \mathcal{U}^0(t-t') \mathcal{\Perturb}_\Enslabel \mathcal{U}^0(t'-t'') \mathcal{\Perturb}_\Enslabel \mathcal{U}^0 (t'') \nonumber\\
 	 &  +\ldots,\nonumber
 \end{align}
 where $\mathcal{U}_\Enslabel^0(t)$ is the Liouville time-evolution superoperator associated with the unperturbed part $\Ham^0_\Enslabel$, and $\mathcal{\Perturb}_\Enslabel \Liouket{\rho} = [\Perturbop_\Enslabel, \Dm]$ is the Livouillian associated with the perturbation $\Perturbop_\Enslabel$. For small perturbations ($\Perturbparam \ll 1$), the above Eq.~\eqref{eq:Perturbation expansion Liouville} can be truncated to a given order in $\Perturbparam$. 
 
To proceed further, we represent the unitary time-evolution superoperator $\mathcal{U}_\Enslabel(t)$ by its dynamical matrix $\Dmatrix_\Enslabel(t)$ and identify the Liouville states $\Liouket{\rho}$ with vectors $\Dmvector$ \cite{Kropf2016a}, such that 
\begin{align}
 \Liouket{\Dmonly_\Enslabel(t)} =\mathcal{U}_\Enslabel (t) \Liouket{\Dmonly(0)}  &\leftrightarrow \Dmvector_\Enslabel(t) = \Dmatrix_\Enslabel(t) \cdot \Dmvector(0)
\end{align}
Concretely, the dynamical matrix $\Dmatrix_\Enslabel(t)$ for a single realisation of the static noise is obtained by expanding $\mathcal{U}_\Enslabel(t)$ in an orthonormal basis and identify the matrix elements. Given $\ens{\Liouket{jk}}$ an orthonormal basis in Liouville space, and $\ens{\ket{j}}$ the common eigenvectors of the unperturbed Hamiltonians $\Ham_\Enslabel^0$, we have
 \begin{align}\label{eq:Perturb F Liouville}
 	\Lioumatelem{jk}{\mathcal{U}_\Enslabel (t)}{rs} = \matelem{j}{\op{U}_\Enslabel(t)}{r}\matelem{s}{{\op{U}}^\dagger_\Enslabel(t)}{k} = \Dmatrix_{jk,rs}^\Enslabel(t).
 \end{align}
The dynamical matrix has double indices $(jk)$ with $j,k = 1,\ldots, d$ ordered as $(11), (12), \ldots (1d), (21),\dots$, with $d$ the dimension of the system.
 We can now expand the evolution operator Eq.~\eqref{eq:Perturbation expansion Liouville} in the basis $\ens{\Liouket{jk}}$ of the unperturbed Hamiltonians, and identify the corresponding dynamical matrix for each order in $\Perturbparam$:
 \begin{align} \label{eq:Pertubative dynamical matrix}
 	\Dmatrix_{\Enslabel}(t) = \Dmatrix_{\Enslabel}^{0}(t)+\Perturbparam\Dmatrix_{\Enslabel}^{1}(t)+\Perturbparam^2\Dmatrix_{\Enslabel}^{2}(t)+\ldots.
 \end{align}
The matrix elements of each order can be evaluated separately (detailed calculations can be found in \cite{Kropf2017}).
 
 The ensemble average of the dynamical matrix, is obtained by taking the weighted integral over all realisations of the noise of the perturbative dynamical matrix, Eq.~\eqref{eq:Pertubative dynamical matrix}, and reads
 \begin{align}
 	\avg{\Dmatrix}(t) &= \int d\Enslabel p_\Enslabel \left[\Dmatrix_\Enslabel^0(t)+\Perturbparam\Dmatrix_\Enslabel^1(t)+\Perturbparam^2\Dmatrix_\Enslabel^2(t)+\dots\right] \nonumber\\&= \avg{\Dmatrix}^0(t)+\Perturbparam \avg{\Dmatrix}^1(t) + \Perturbparam^2 \avg{\Dmatrix}^2(t) +\ldots.\label{eq:Perturb average F expansion}
 \end{align}
 Note that the ensemble-averaged dynamics, Eq.~\eqref{eq:EnsAvgDyn}, are obtained from $\vec{\avg{\rho}}(t) = \avg{\Dmatrix}(t) \cdot\vec{\rho} (0)$, which in terms of density matrix elements reads $\matelem{j}{\Avgdm(t)}{k} = \sum_{r,s=1}^d \avg{\Dmatrix}_{jk,rs}(t)\matelem{r}{\Avgdm(t)}{s}$.
 
To derive a corresponding perturbative disorder master equation, we follow the general procedure introduced in \cite{Kropf2017}, which is analogous to the time-convolutionless expansion (TCL) approach from the theory of open quantum system \cite{Breuer2002}. The first step is to compute the matrix $\avg\QMEmatrix(t) = \dot{\avg{\Dmatrix}}(t) \cdot \avg{\Dmatrix}^{-1}(t)$ to obtain a time-convolutionless form of the generator, i.e. $\dot{\vec{\avg{\rho}}} = \avg\QMEmatrix(t) \cdot \vec{\avg{\rho}}$. The time derivative $\dot{\avg{\Dmatrix}}(t)$ is directly obtained by taking the time derivative of Eq.~\eqref{eq:Perturb average F expansion}.
 Computing the inverse $\avg{\Dmatrix}^{-1}(t)$ analytically is, in general, not possible. We therefore make use of the perturbative nature of the dynamics and express the inverse with the help of the Neumann series  \cite{Neumann1877} as $\avg\Dmatrix^{-1}(t) = \sum_{n=0}^\infty \left(\Id -{\avg{\Dmatrix^0}}^{-1}(t)\avg\Dmatrix(t)\right)^n {\avg{\Dmatrix^0}}^{-1}(t)$ (c.f. Appendix \ref{app:Neumann}). 

The expression for $\avg{\QMEmatrix}(t) =\dot{\avg{\Dmatrix}}(t)\cdot \avg{\Dmatrix}^{-1}(t) $ then yields 
\begin{align}\label{eq:QmatrixDef}
	\avg{\QMEmatrix}(t) =&\left(\sum_{n=0}^\infty \Perturbparam^n \dot{\avg{\Dmatrix^n}}(t)\right) \\ &\cdot \sum_{m=0}^\infty \left(-\sum_{k=1}^\infty \Perturbparam^k \avg{\Dmatrix^0}^{-1}(t) \avg{\Dmatrix^k}(t) \right)^m\avg{\Dmatrix^0}^{-1}(t).\nonumber
\end{align}
The matrix $\avg{\QMEmatrix}(t)$ fully characterizes the perturbative master equation for the ensemble-averaged dynamics, Eq.~\eqref{eq:EnsAvgDyn}, and can be truncated at the appropriate order in the perturbation parameter $\Perturbparam$. If desired, a Lindblad form of the master equation can be obtained by expanding $\QMEmatrix(t)$ in a basis of traceless, Hermitian operators \cite{Andersson2007,Breuer2002,Kropf2016a}. Alternatively, one can express the master equation in terms of the density matrix components
\begin{align}
	\dot{\Avgdm}(t) = \sum_{j,k=1}^d \ketbra{j}{k} \sum_{r,s=1}^d \avg{\QMEmatrix}_{jk,rs}(t)\matelem{r}{\Avgdm(t)}{s},
\end{align}
and then collect and rearrange the terms in a suitable way.
 
 \section{Perturbative master equation for diagonal static noise/disorder}\label{sec:PME for on side disorder}
In this section we derive the explicit form of the perturbative master equation to first order for $d$-dimensional systems with diagonal (spectral) static noise and a weaker, noise- and time-independent perturbation. More precisely, the random Hamiltonians read
\begin{align}\label{eq:Anderson Ham}
	\Ham_\Enslabel = \Ham_\Enslabel^0 + \Perturbparam \Perturbop
\end{align}
where $\Perturbparam \ll |\Eno_{\Enslabel,j}-\Eno_{\Enslabel,k}|$ for all $j,k$ such that $V_{jk}\neq 0$. In other words, the coupling potential between two unperturbed eigenstates is much smaller than their energy difference. The system can thus be treated with non-degenerate perturbation theory \cite{Sakurai2011}. 

Since we assume the perturbation to be independent of the noise ($\op{\Perturb}_\Enslabel = \op{\Perturb}$), the eigenvalues $\Eno_{\Enslabel,j}$ of the unperturbed Hamiltonian $\Ham_\Enslabel^0$ are the only random variables. Indeed, we remind the reader we assume the eigenvectors $\ens{\ket{j}}$ of $\Ham_\Enslabel^0$ to be independent of the  noise realisation (on the contrary the eigenvectors of $\Ham_\Enslabel$ are obviously modified by the noise). If the eigenvectors of $\Ham_\Enslabel^0$ would also be subject to noise, one would have to expand the dynamical matrix, Eq.~\eqref{eq:Pertubative dynamical matrix}, in a noise-independent basis before taking the ensemble average. This does not change the derivation procedure but the computations would, in general, become more intricate. Furthermore, we define, without loss of generality, that in the eigenbasis of the unperturbed Hamiltonian $\Ham_\Enslabel^0$, the diagonal elements of the perturbation vanish, $\matelem{j}{\Perturbop}{j}=0$ ; possible contributions of $\Perturbop$ to the diagonal can be absorbed into the definition of the ensemble-averaged value of the unperturbed Hamiltonian $\Avgham^{0} = \int d\Enslabel \, p_\Enslabel \Ham_\Enslabel^0$. 

We parametrize the random Hamiltonian ensemble as
\begin{align}\label{eq:Perturb ensemble}
	\ens{\Ham_{\vec{\Enslabel}} = \Ham^0_{\vec{\Enslabel}} + \Perturbparam \Perturbop  \;\;,\;\; p_{\vec{\Enslabel}} }
\end{align}
with $p_{\vec{\Enslabel}} = p(\Enslabel_1,\Enslabel_2,\ldots,\Enslabel_\Dim)$ the joint probability density distribution of the dimensionless variables $\Enslabel_j$ that characterize the eigenvalues $\Eno_{\Enslabel,j}= \Eavg_j + \hbar \Escale \Enslabel_j$ ($j=1,2,\dots,\Dim$) of $\Ham^0_\Enslabel$, and where $\hbar \Escale$ is the reference energy scale and $\Eavg_j$ the noise-free part.

Interestingly, and as will become clear in the next section, the first order correction to the master equation matrix $\avg\QMEmatrix(t)$ is enough to capture the main features of the ensemble-averaged dynamics at all times for random ensembles of the form of Eq.~\eqref{eq:Perturb ensemble}. The first order in $\Perturbparam$ master equation can be expressed as (for a detailed derivation see App.~\ref{sec:Perturb weak PME derivation})
\begin{align*}
			&\dot{\Avgdm}(t) = \sum_{j,k=1}^\Dim \PrateO_{jk}(t) \Projectionop_{jj} \Avgdm(t) \Projectionop_{kk}-\Perturbparam\frac{i}{\hbar} \commutator{\Perturbop}{\Avgdm(t)} \\
	& +\frac{\Perturbparam}{\hbar} \sum_\subsum{j,k=1}{j\neq k}^\Dim \PrateA_{jk}(t) \Perturb_{jk} \left[\Projectionop_{jk}\Avgdm(t)\Projectionop_{kk}-\Projectionop_{jj}\Avgdm(t)\Projectionop_{jk}\right] \numberthis \label{eq:Perturb weak QME1}\\
		& +\frac{\Perturbparam}{\hbar} \sum_\subsum{j,k,r=1}{j\neq k\neq r}^\Dim \PrateB_{jkrj}(t) \Perturb_{jr} \Projectionop_{jr}\Avgdm(t)\Projectionop_{kk} + \PrateB_{jkrj}^*(t)\Perturb_{rj} \Projectionop_{kk}\Avgdm(t)\Projectionop_{rj}
\end{align*}
with the diadic operators $\Projectionop_{jk} := \ketbra{j}{k}$. The (time-dependent) rates in the equation above are functions of the phase factors
\begin{align}\label{eq:Perturb avg phases}
 	\Phase_{jk}(t) & := e^{-\frac{it}{\hbar}(\Eno_{\Enslabel,j}-\Eno_{\Enslabel,k})} = e^{-\frac{it}{\hbar}(\Eavg_j-\Eavg_k)} e^{-it\Escale(\Enslabel_j-\Enslabel_k)},
 \end{align}
and their average $\avg{\Phase}_{jk}(t) = \int d \vec\Enslabel p_{\vec{\Enslabel}} \Phase_{jk}(t)$. The rates are defined as
\begin{align*}
 \PrateO_{jk}(t) &\eqdef \frac{d}{dt} \ln \left[\avg{\Phase}_{jk}(t)\right] \numberthis\label{eq:Perturb PrateO def} \\
 \PrateA_{jk}(t) &\eqdef i\left(1-\avg{\Phase}_{jk}(t) + \PrateO_{jk}(t)\int_0^t dt' \avg{\Phase}_{jk}(t')\right) \\ \nonumber &= -\PrateA_{kj}^*(t) \numberthis \label{eq:Perturb PrateA def}, \\
	\PrateB_{jkrj}(t) &\eqdef i\Biggl[\frac{\dot{\avg{\Phase}}_{jk}(t)}{\avg{\Phase}_{jk}(t)\avg{\Phase}_{rk}(t)} \avg{\Phase_{jk}(t)\int_0^t dt' \,\Phase_{rj}(t')} \\
	&\quad - \frac{1}{\avg{\Phase}_{rk}(t)} \left( \avg{\dot{\Phase}_{jk}(t)\int_0^t dt' \,\Phase_{rj}(t')}\right)\Biggr]\numberthis \label{eq:Perturb PrateB def} \\
	=&-\PrateB_{kjjr}^*(t).  
\end{align*}
Here the symmetry $\PrateB_{jkrj} = -\PrateB_{kjjr}^*$ is obtained by permuting the first two indices, $jk\rightarrow kj$, and the last two indices, $rj\rightarrow jr$, and reflects the Hermiticity of the density matrix (the dynamical matrix $\Dmatrix(t)$ has the same symmetry as $\PrateB(t)$).
The first-order in $\Perturbparam$ part of the master equation, Eq.~\eqref{eq:Perturb weak QME1}, is manifestly separated into a fully coherent contribution from the perturbation potential $\Perturbop$ (second term, first line), and additional incoherent terms with decoherence rates $\propto \PrateA(t) $ and $\propto\PrateB(t)$ (second and third lines).

In order to better understand the dynamics arising from the master equation, Eq.~\eqref{eq:Perturb weak QME1}, we express it in terms of the density matrix elements. For the coherences (off-diagonal elements $j\neq k$) we have
\begin{align*}
	&\matelem{j}{\dot{\Avgdm}(t)}{k} = \PrateO_{jk}(t)\matelem{j}{\Avgdm(t)}{k} \\&-i\frac{\Perturbparam}{\hbar} \sum_{r=1}^\Dim V_{jr}\matelem{r}{\Avgdm}{k}-\matelem{j}{\Avgdm}{r}V_{rk}\\ &+\frac{\Perturbparam}{\hbar}\left[\PrateA_{jk}(t)-i\right]\Perturb_{jk} \left[\matelem{k}{\Avgdm(t)}{k} - \matelem{j}{\Avgdm(t)}{j}\right] \numberthis \label{eq:Perturb weak coherences QME}\\&+\frac{\Perturbparam}{\hbar} \sum_{\mathclap{\substack{r=1\\ r\neq j\neq k}}}^\Dim \left[\PrateB_{jkrj}(t)-i\right]\Perturb_{jr}\matelem{r}{\Avgdm(t)}{k}\\&\quad\quad\quad\quad\quad\quad\quad -\left[ \PrateB_{jkrj}^*(t)+i\right]\Perturb_{rk}\matelem{j}{\Avgdm(t)}{r}.
\end{align*}
 The term proportional to $\PrateO_{jk}(t)$ describes the effects of pure diagonal disorder ($\Perturbparam = 0$) \cite{Kropf2016a}, and leads to time-dependent dephasing in the eigenbasis of $\avg{\Ham^0}$. All other terms arise as a consequence of the perturbation potential $\Perturbop$. The rate $\PrateA_{jk}(t)$ is associated with the dynamical coupling of the coherences to the population differences, whereas the rate $\PrateB_{jkrj}(t)$ governs the second-order coupling of the coherences to the other off-diagonal terms of the density matrix.

At the same time, for the populations (diagonal terms $j=k$) we have
\begin{align}\label{eq:Perturb weak populations QME}
	\matelem{j}{\dot{\Avgdm}(t)}{j} =&-i\frac{\Perturbparam}{\hbar} \sum_\subsum{r=1}{r\neq j}^\Dim \Perturb_{jr}  \matelem{r}{\Avgdm(t)}{j} - \matelem{j}{\Avgdm(t)}{r}\Perturb_{rj}.  
\end{align}
Hence the populations do evolve in time (as opposed to the case when $\Perturbparam = 0$), but since Eq.~\eqref{eq:Perturb weak populations QME} does not contain any noise-dependent decoherence term, their evolution is not directly affected by the noise. They only indirectly feel the noise through their coupling to the coherences via the perturbative potential $\Perturbop$.

The ensemble-averaging for ensembles of the type \eqref{eq:Perturb ensemble} thus leads to complex, in general time-dependent, dephasing (decoherence) processes.  

\subsection{Statistical interpretation}

The ensemble-averaged phase factors $\avg\Phase_{jk}(t)$ that characterize the decoherence rates, Eqs.~\eqref{eq:Perturb PrateO def}-\eqref{eq:Perturb PrateB def}, can be expressed in terms of the complex conjugate of the characteristic function \cite{Lukacs1972}
\begin{align}\label{eq:CharacFunction}
 \phi_{jk}(\Escale t) = \mathbb{E}\left[e^{i t\Escale\Delta_{jk}}\right]
\end{align}
of the probability density distribution $q_{jk}(\Delta_{jk}) :=\int d\vec{\Enslabel} p_{\vec{\Enslabel}}\delta(\Delta_{jk}-(\Enslabel_j-\Enslabel_k))$ of the difference of pairs of random variables $\Delta_{jk} :=\Enslabel_j - \Enslabel_k$ as
\begin{align}
	\avg\Phase_{jk}(t) &=  e^{-\frac{it}{\hbar}(\Eavg_j-\Eavg_k)}\Charact_{jk}(\Escale t) \\
	&= e^{-\frac{it}{\hbar}(\Eavg_j-\Eavg_k)}\int d\Delta_{jk} q_{jk}(\Delta_{jk})e^{-it\Escale \Delta_{jk}}. 
\end{align}
Hence, the time-dependent properties of the ensemble-averaged dynamics can be traced-back to the properties of the characteristic functions of the pair-wise eigenvalues differences distributions. We remark that higher order moments (involving more than two eigenvalues) are necessary if one  considers contributions beyond the first-order in the perturbation. 

Conveniently, the characteristic function can be described in terms of its cumulants \cite{Kenney1947} which yield a direct physical interpretation\cite{Kropf2016} -- the odd cumulants capture the (a)symmetry of the distribution, and contribute to the coherent (Hamiltonian) part of the master equation, while the even cumulants describe the width of the distribution (or strength of the noise), and characterize the decoherence rates Eqs.~\eqref{eq:PrateOLargeTimeS}-\eqref{eq:PrateBLargeTimeS}, i.e., the time-dependence and speed of the dephasing process.

Note that while the characteristic function describes the time-dependence of the decoherence process, the structure of the Hamiltonian, and in particular the nature of the coupling $\Perturbop$, determines the genre of the decoherence \cite{Kropf2016a}. Indeed, the coupling potential determines the Lindblad operators of the master equation \eqref{eq:Perturb weak QME1}, i.e., the type of dephasing/decoherence dynamics. This is consistent with findings in \cite{Kropf2016a} that the structure of the disorder defines the form of the Lindblad operators. 
 
\subsection{Range of validity}\label{sec:Validity}
In the derivation of the general perturbative master equation (c.f. Section \ref{sec:PME derivation}) we assume the convergences of the Neumann series used to expand the inverse of the ensemble-average dynamical matrix $\avg{\Dmatrix}$, i.e.,  $\limit{n}{\infty} \left(\Id - {\avg{\Dmatrix^0}}^{-1}\avg{\Dmatrix}\right)^n =0$. In other words, we assume that non-degenerate perturbation theory applies to most realizations of the noise, which requires $\Perturbparam V_{jk} \ll \Eno_{\Enslabel,j} - \Eno_{\Enslabel,k}$ for most Hamiltonians $\Ham_\Enslabel$ in the random ensemble Eq.~\eqref{eq:Perturb ensemble}. Hence, the eigenenergy distribution shows level-repulsion for all state coupled by $\Perturbop$. This is for instance satisfied for chaotic quantum systems, which by definition exhibit level repulsion\cite{Haake2010,Tomsovic1994,Wigner1951}. A multitude of other systems can also be treated within the scope of non-degenerate perturbation theory, such as the hopping of an excitation in a one-dimensional chain with an electric potential ladder \cite{Kropf2019}, or in a molecular network \cite{Adolphs2007}.

As we will show with examples in Section \ref{sec:LongRange} below, if the above conditions are satisfied, the perturbative master equation, Eq.~\eqref{eq:Perturb weak QME1}, captures the ensemble-average dynamics not only on transient time scales, but also on asymptotic time scales (up to systematic errors). This can be understood as a consequence of the interplay between the accumulated errors in the phases from the perturbative expansion, and the ensemble-averaging induced dephasing. On the one hand, the finite perturbative expansion leads to an approximation of the exact phases and eigenvectors of the system. In a closed system these errors in the phases accumulate over time, and eventually any phase relation between the perturbative dynamics and the actual dynamics is lost at sufficiently long times. On the other hand, the ensemble-averaging induced dephasing leads to an overall diminution of the phases and keeps the accumulated error at a finite value. Hence, as long as the dephasing process is fast enough to keep the error from perturbation theory finite, the  effective non-unitary dynamics arising from the corresponding perturbative master equation only incur a systematic error. 

When the conditions for non-degenerate perturbation theory are not satisfied, the perturbative master equation captures the dynamics at least on short time-scales in the sense discussed in Refs.~\cite{Kropf2016a,Kropf2017a,Gneiting2016}. Indeed, the first-order in time approximation of Eq.~\eqref{eq:Perturb weak QME1} yields
\begin{align*}
	&\dot{\Avgdm}(t)= -\frac{i}{\hbar}\commutator{\Avgham}{\Avgdm(t)} \numberthis \label{eq:Perturb weak ST QME}\\&+2\frac{t}{\hbar^2}\sum_{j,k=1}^\Dim\left(\avg{\Eno_j}\avg{\Eno_k}-\avg{\Eno_j\Eno_k}\right)\Projectionop_{jj} \Avgdm(t) \Projectionop_{kk}.
\end{align*}
This equation corresponds to the short-time master equation \cite{Kropf2016a} 
\begin{align*}
	\dot{\Avgdm} = -\frac{i}{\hbar}\left[\Avgham,\Avgdm(t)\right] + \int d\Enslabel p_\Enslabel \left(\Lop_\Enslabel\Avgdm\Lop_\Enslabel -\frac{1}{2}\{\Lop_\Enslabel^\dagger\Lop_\Enslabel,\Avgdm\}\right)
\end{align*}
with $\Lop_\Enslabel = (\Ham_\Enslabel - \Avgham)/(\hbar\Escale)$, which captures the effective ensemble-averaged dynamics of any random Hamiltonian ensemble at times $t \ll 1/\Escale$, which corresponds to the Gaussian decay regime \cite{JulianeAndMe}. 

\subsection{Symmetric noise distributions and decoherence rates} \label{sec:PhysicalNoise}
Physical noise distributions $q_{ij}(\Delta_{jk})$ are typically symmetric \cite{Weiss1999}, such as, e.g., Gaussian, Lorentzian or uniform distributions. The non-degenerate constraint $\Eno_{\Enslabel,j}-\Eno_{\Enslabel,k} = (\Eavg_j-\Eavg_k) + \hbar\Escale\Delta_{jk} \gg \Perturbparam$ for most realization of the noise then implies that the distribution must be sufficiently peaked ; if the distribution is too wide, the probability for having degenerate levels could become too high, restricting the validity of the perturbative master equation to short times (Gaussian decay regime), c.f. Section \ref{sec:Validity}. The decoherence rates Eqs.~\eqref{eq:Perturb PrateO def}-\eqref{eq:Perturb PrateB def} can then be expressed in terms of the real part (even cumulants) of the characteristic function  $\Charact_{jk}(t)$ and the central values (first cumulant) $ \avg{\Eno_j} = \Eavg_j + \hbar\Escale\avg{\Enslabel}_j$ (c.f. Appendix \ref{app:Rates}),
\begin{align}\label{eq:PrateOLargeTimeS}
	&\PrateO_{jk} = -i[\avg{\Eno_j} /\hbar- \avg{\Eno_k}/\hbar] + \real{\frac{\dot{\Charact}_{jk}}{\Charact_{jk}}} \\
		&\PrateA_{jk}(t) \approx \left[1-e^{-\frac{it}{\hbar}(\Eavg_j-\Eavg_k)}\Charact_{jk}(\Escale t)\right] \frac{ \real{\PrateO_{jk}(t)}}{\avg{\Eno_j} /\hbar- \avg{\Eno_k}/\hbar} \label{eq:PrateALargeTimeS}\\
		&\PrateB_{jkrj}(t) \approx \frac{\left[\real{\PrateO_{jk}(t)}-\real{\PrateO_{rk}(t)}\right]}{\avg{\Eno_j} /\hbar- \avg{\Eno_k}/\hbar}.\label{eq:PrateBLargeTimeS}
 \end{align}
The zeroth-order function $\PrateO_{jk}$ separates into an imaginary coherent (energy shift) contribution and a real-valued, time-dependent decoherence rate. The first-order time-dependent decoherence rate $\PrateB_{jkrj}(t)$ is real-valued, whereas $\PrateB_{jk}(t)$ is modulated by a fast decaying, complex oscillating function $ 1-e^{-\frac{it}{\hbar}(\Eavg_j-\Eavg_k)}\Charact_{jk}(\Escale t)$. Note that by the Riemann-Lebesgue lemma \cite{Lebesgue1906,Riemann1867} the characteristic function vanishes at large times, i.e., $\Charact_{jk}(t) \overset{t\rightarrow \infty}{\longrightarrow} 0$.

We remark that if the distribution $q_{jk}$ is not symmetric, such as, e.g., a general Lévy distribution, one obtains time-dependent energy shifts, i.e., $\imag{\PrateO_{jk}} = \imag{\PrateO_{jk}}(t)$ \cite{Kropf2016a}. In addition, the energy terms $[\avg{\Eno_j} /\hbar- \avg{\Eno_k}/\hbar]$ in the decoherence rates, Eqs.~\eqref{eq:PrateALargeTimeS} and \eqref{eq:PrateBLargeTimeS}, must be replaced by $-\imag{\PrateO_{jk}}(t)$.

\subsection{Asymptotic state}
Given the decoherence rates \eqref{eq:PrateOLargeTimeS}-\eqref{eq:PrateBLargeTimeS}, the asymptotic state of the perturbative master equation can be expressed in terms of the ensemble-averaged Hamiltonian $\Avgham$. Indeed, the latter is at large times in the kernel of the perturbative master equation: upon insertion of $\Avgham = \Avgham^0 + \Perturbparam V = \sum_j (\Eavg_j+\hbar\Escale \avg{\Enslabel}_j)\ketbra{j}{j} + \Perturbparam \op{\Perturb}$ into Eq.~\eqref{eq:Perturb weak QME1}, and using the rates \eqref{eq:PrateOLargeTimeS}-\eqref{eq:PrateBLargeTimeS}, we see that the first commutator vanishes, the second and third sum cancel each other for $t\gg 0$ because $\Charact_{jk}(t) \rightarrow 0$, and the fourth sum is of order $\Perturbparam ^ 2$ (note that the fourth sum vanishes exactly for identically and independently distributed random variables (i.i.d.) because in this case $\PrateB_{jkrj}(t) \equiv 0$). The ensemble-average Hamiltonian thus characterizes the asymptotic states up to first order in $\Perturbparam$. Note that this does not mean that there is a unique asymptotic state; rather, as we will show in the examples below, the perturbative master equation describes asymptotically a dephasing process in the eigenbasis of $\Avgham = \sum_n E_n \ketbra{n}{n}$. Hence, an initial state $\Dm_0$ will asymptotically be projected onto the basis $\left\{\ket{n}\right\}$,
\begin{align}\label{eq:AsymptState}
	\Dm_0 \rightarrow \Avgdm_\infty \approx \sum_n \ketbra{n}{n}\Dm_0\ketbra{n}{n}.
\end{align}
As will become clear with the examples in Section \ref{sec:LongRange}, this means that the coherences (off-diagonal elements $\avg{\rho}_{jk}(0)$) of the initial state expressed in the eigenbasis of $\Ham_0$, which are coupled by the perturbation $\Perturb_{jk}$, are partially protected from the averaging-induced dephasing. 

\section{Applications}\label{sec:LongRange}

As applications of the perturbative master equation for diagonal noise we consider three examples. (A) First, we study the dynamics of a single qubit which allows for a geometric intuition in terms of Bloch vectors of the dephasing dynamics arising from the ensemble-average. (B) Second, we consider a one-dimensional lattice model with uncorrelated on-site disorder and study the effect of short and long-range interactions on the coherences in the asymptotic state. (C) Third, we investigate the effect of strong correlations on the effective dynamics of the coherences in a many-particle boson model.

\subsection{Geometrical interpretation: Qubit with Gaussian energy disitribution}\label{sec:Perturb single Qubit}
We begin with two-level systems ($\Dim=2$) with static noise in the energy difference, that physically, for instance, may describe an ensemble of spins $1/2$ precessing in a static, spatially inhomogeneous magnetic field. It is known that this noise leads to the inhomogeneous broadening of the line-width in spectroscopy experiments, and characterizes the decay of the total magnetization on the timescale $T_2^*$ \cite{Slichter1990}. While the effect of the ensemble-average decoherence can in this case be cancelled out by inverting the spin-precession direction with a spin-echo sequence \cite{Slichter1990}, the latter may also be subject to noise making the inversion incomplete \cite{Fine2014}. Thus, understanding the exact dynamical role of the static noise is of relevance. 

The static noise ensemble is parametrized as
\begin{align}\label{eq:Perturb Qubit Ham}
	\ens{\Ham_\Enslabel = \hbar \Escale\left( \frac{\Enslabel}{2} \op\sigma_z + \Perturbparam \op\sigma_x\right) = \frac{\hbar \Escale}{2}\begin{pmatrix}
	\Enslabel & 2\Perturbparam \\ 2\Perturbparam & -\Enslabel
	\end{pmatrix}\;\; , \;\; p_{\vec{\Enslabel}} = p(\Enslabel)},
\end{align}
with $\hbar\Escale$ the reference energy, $\Enslabel$ a dimensionless random variable, and $\Perturbparam$ a dimensionless perturbation parameter. The unperturbed Hamiltonian $H_\Enslabel^0 = \hbar \Escale\Enslabel \sigma_z /2$ is diagonal, with eigenvalues $\Eno_{\Enslabel,1} =  \hbar \Escale \Enslabel /2$ and $\Eno_{\Enslabel,2} = -\hbar \Escale \Enslabel /2$ (here $\Eavg_{1,2}=0$), and the perturbed part is $\Perturbparam\Perturbop = \Perturbparam \hbar \Escale\sigma_x$. Referring to the physical picture of spins in a static magnetic field, we assume that the field has a static random amplitude along the z-axis sampled from the probability distribution $p(\Enslabel)$, and a small, constant component along the x-axis proportional to $\Perturbparam$. Hence, in total, not only the amplitude of the magnetic field varies from realization to realization, but also the orientation.

We consider a generic Gaussian noise distribution, with average value $\Enslabel_0 \in \mathbb{R}$ and variance $\Pwidth >0$,
\begin{align}\label{eq:GaussianDistr}
	p_\textrm{Ga}(\Enslabel) = \frac{1}{\Pwidth\sqrt{2\pi}}e^{-\frac{(\Enslabel-\Enslabel_0)^2}{2\Pwidth^2}},
\end{align}
which implies $\avg{\Eno_{1}} = \hbar\Escale\Enslabel_0/2$ and $\avg{\Eno_{2}} = -\hbar\Escale\Enslabel_0/2$.
The non-degeneracy perturbation condition then requires $\abs{\Enslabel_0} \gtrsim \Pwidth + \Perturbparam$, i.e., the noisy magnetic field along the z-axis is always much larger than the one along the x-axis.

\begin{figure}
	\includegraphics[width = \linewidth]{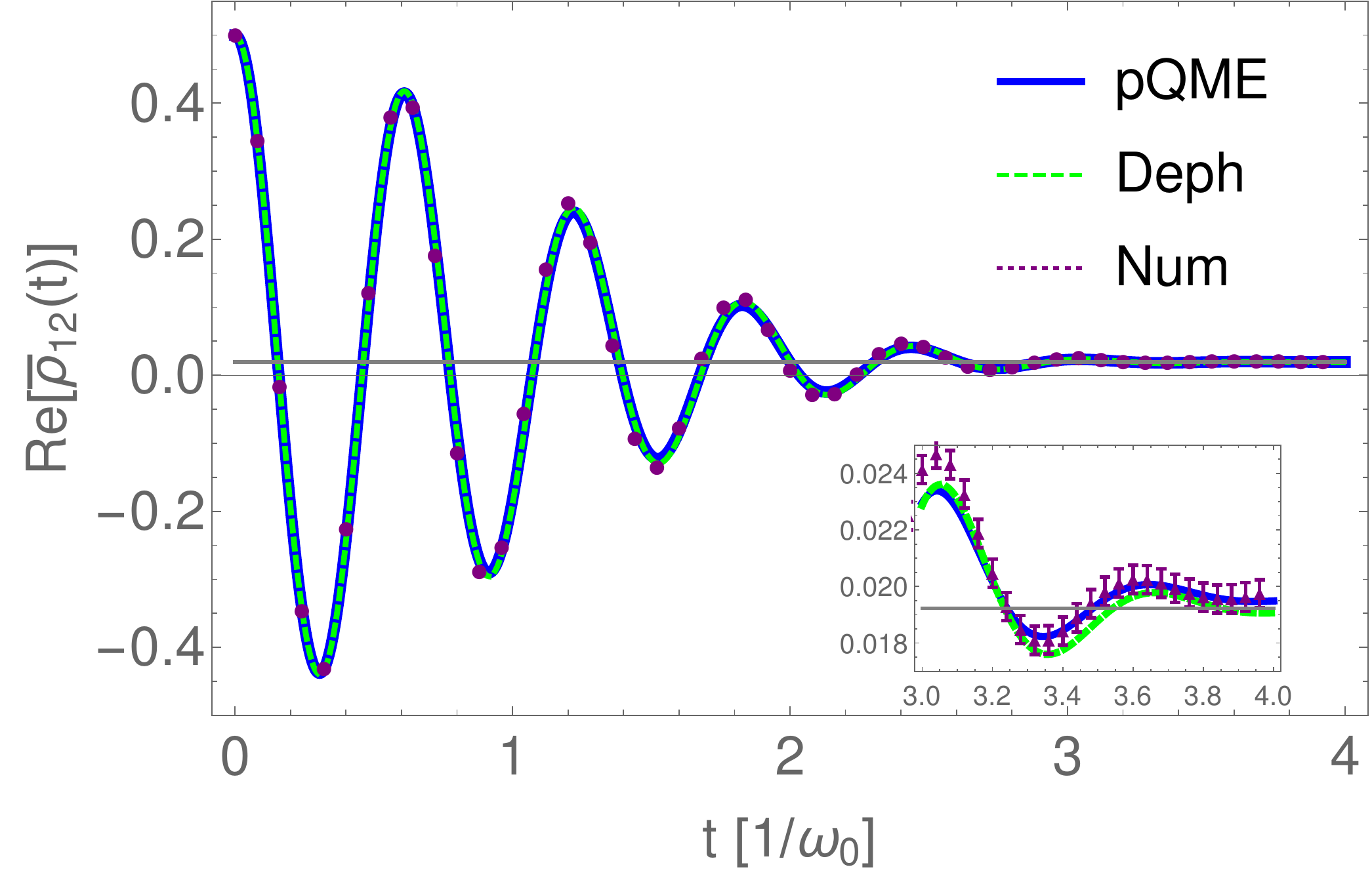}
	\caption{Decay of the coherences (x-component) of a qubit described by the noisy ensemble Eq.~\eqref{eq:Perturb Qubit Ham} obtained from numerical integration of the perturbative quantum master equation Eqs.~\eqref{eq:Perturb Qubit rho12dot}\eqref{eq:Perturb Qubit rho11dot} (blue full), from the pure dephasing approximation Eq.~\eqref{eq:PureDeph} (green, dashed), and from direct numerical averaging (purple dots, $10^6$ realizations). The parameters are set to $\Enslabel_0 = 10,\Pwidth=1,\Perturbparam =1$ and $\rho_0 = 1/2\Id + 1/2\sigma_x$. Inset: The coherences $\real{\avg{\rho}_{12}}$ do not vanish asymptotically and converge to Eq.~\eqref{eq:AsymptState} (gray horizontal line).}
	\label{fig:FID}
\end{figure}

Since there are only two levels (i.e., $j,k = 1,2$) characterized by one single noise parameter $\Enslabel$, the decoherence rates $\PrateO_{jk}(t)$, $\PrateA_{jk}(t)$ and $\PrateB_{jkrj}(t)$  are fully characterized by $\Charact_{12}(\Escale t) =  \int d\Enslabel p(\Enslabel) e^{-it\Escale\Enslabel} = \textrm{exp}[-i\Escale\Enslabel_0 t-1/2\Escale^2\Pwidth^2 t^2]$, Eq.~\eqref{eq:CharacFunction}. From Eqs.~\eqref{eq:PrateOLargeTimeS}-\eqref{eq:PrateBLargeTimeS}, we obtain
\begin{align}\label{eq:QubitPrateO}
	\PrateO_{12}(t)=\PrateO(t) = -i\Escale\Enslabel_0 - \Escale^2\Pwidth ^2 t, 
\end{align}
\begin{align}\label{eq:QubitPrateA}
	\PrateA_{12}(t) = \PrateA(t) \approx -\frac{\Escale\Pwidth^2 t}{\Enslabel_0}\left(1-e^{-i\Escale\Enslabel_0-\frac{\Escale^2\Pwidth^2}{2} t^2}\right)
\end{align}
and $\PrateB_{jkrj}(t) = 0$ for $\Dim <3$ because it requires three distinct indices $j,k,r$. It follows that the coherences, Eq.~\eqref{eq:Perturb weak coherences QME}, evolve according to
\begin{align}\label{eq:Perturb Qubit rho12dot}
	\matelem{1}{\dot{\Avgdm}(t)}{2} &= \left(-i\Escale\Enslabel_0 - \Escale^2\Pwidth ^2 t\right)\matelem{1}{\Avgdm(t)}{2} \\
	&+\Perturbparam\left(\PrateA(t)-i\right)\left[\matelem{2}{\Avgdm(t)}{2} -\matelem{1}{\Avgdm(t)}{1} \right],\nonumber
\end{align}
and, by definition, $\matelem{2}{\dot{\Avgdm}(t)}{1}  =\matelem{1}{\dot{\Avgdm}(t)}{2}^* $. The populations, Eq.~\eqref{eq:Perturb weak populations QME}, satisfy the equations 
\begin{align}\label{eq:Perturb Qubit rho11dot}
	\matelem{1}{\dot{\Avgdm}(t)}{1} &=  i \Perturbparam\Escale \left[\matelem{1}{\Avgdm(t)}{2}-\matelem{2}{\Avgdm(t)}{1}\right],
\end{align}
and, $\matelem{2}{\dot{\Avgdm}(t)}{2} = -\matelem{1}{\dot{\Avgdm}(t)}{1}$.
Thus, the diagonal elements of the ensemble-averaged state evolve coherently under the action of coupling potential $\Perturbop$, while the off-diagonal elements evolve coherently with $\Avgham$, and decay incoherently with a time-dependent rate $\Escale^2\Pwidth^2 t$.

In Fig.~\ref{fig:FID} we show the time-evolution of the coherences for an initial state $\rho_0=1/2(\Id+\sigma_x)$ polarized in the $x$-direction with $\Enslabel_0 = 10$, $\Pwidth =1$ and $\Perturbparam = 1$, which decay to a the non-vanishing value obtained from Eq.~\eqref{eq:AsymptState} (see inset of Fig.~\ref{fig:FID}). It is also shown that the first order master equation well corresponds to the direct numerical averaging of the dynamics.

This scenario could be measured in a free-induction-decay (FID) experiment with nuclear spins \cite{Slichter1990}. After the $\pi/2$-pulse used to flip the magnetization in the transverse plane has been applied, the weak radio-frequency transverse field is not turned off, but only turned  off-resonance. Then, in the rotating frame and considering Gaussian distributed spatial inhomogeneity of the polarization field, we obtain a Hamiltonian ensemble as described in Eq.\eqref{eq:Perturb Qubit Ham}.
The FID decay of the magnetization $M_{x}$ on the $T_2^*$ time-scale in the rotating frame then is proportional to $\tr{\rho \sigma_x} = \real{\Avgdm_{12}}$, and should not decay completely due to the presence of the transverse off-resonance field, c.f. Fig.~\ref{fig:FID}. A similar protection of coherences from dipole-dipole interaction $T_2$ decay by anomalous resonance conditions for the transverse field was discussed in \cite{Kropf2012,Kropf2016}.

The effective dynamics, Eqs.~\eqref{eq:Perturb Qubit rho12dot},\eqref{eq:Perturb Qubit rho11dot}, are best understood by expressing the master equation \eqref{eq:Perturb weak QME1} in the eigenbasis of $\Avgham$ by applying the rotation $\op{R}=e^{-i\theta/2\op\sigma_y}$ to all operators, i.e., $\op{O}\rightarrow \op{O}_r = \op{R}.\op{O}.\op{R}^\dagger$ with $\theta = \arctan[\Enslabel_0 / (2\Perturbparam)]$ the angle between $\Avgham$ and the z-axis. Neglecting the short-time contributions to the rates ($\PrateA(t) \approx -\Escale \Pwidth^2 t / \Enslabel_0$) and the terms proportional to $\Perturbparam \Pwidth^2/\Enslabel_0$ (since $\abs{\Enslabel_0} \gg \Pwidth + \Perturbparam$), we to obtain a pure dephasing equation
 
\begin{figure}
	\includegraphics[width = 0.5\linewidth]{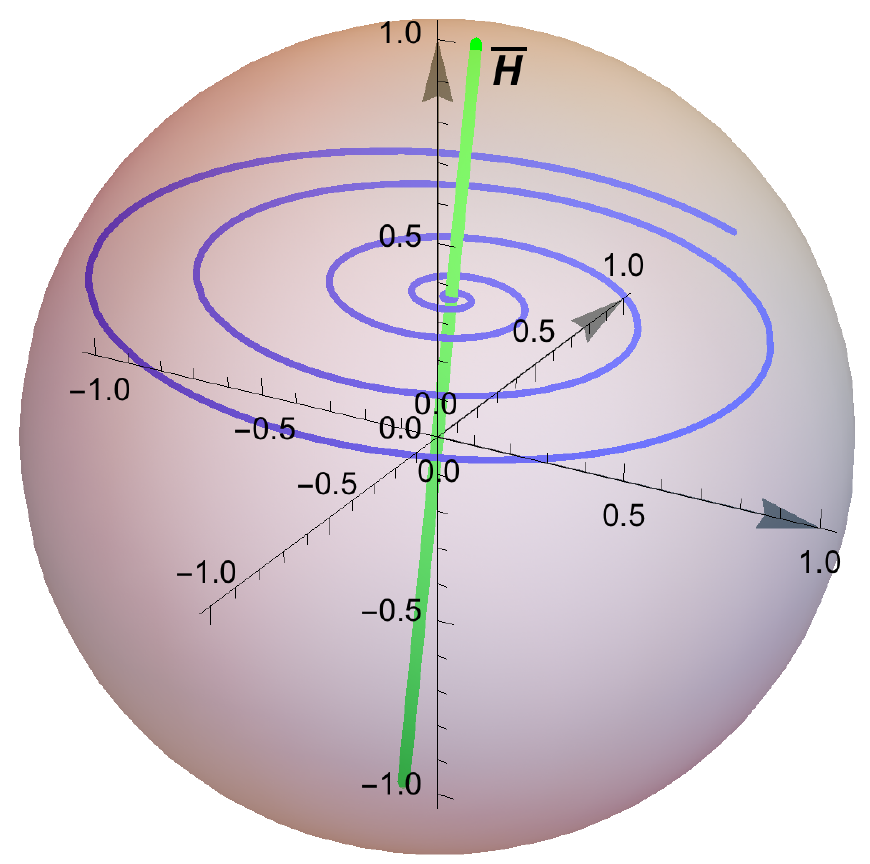}
	\caption{Illustration in the Bloch sphere of the ensemble-averagind induced dephasing process (blue line) Eqs.\eqref{eq:Perturb Qubit rho12dot}\eqref{eq:Perturb Qubit rho11dot} from static diagonal noise. The asymptotic state is the projection of the initial state onto the average Hamiltonian $\Avgham$ (green line) and has non-vanishing coherences (x and y components). For visualization purposes the initial state is different from Fig.~\ref{fig:FID}.}
	\label{fig:QubitAsympt}
\end{figure} 
\begin{align}\label{eq:PureDephQubitQME}
	\dot{\Avgdm}_r &= -\frac{i}{\hbar} [\Avgham_r,\Avgdm_r] \\& +\frac{1}{2}\Escale^2\Pwidth^2 t \left(\op\sigma_z^r\Avgdm_r\op{\sigma_z^r}-\frac{1}{2}\{\op\sigma_z^r\op\sigma_z^r,\Avgdm_r\}\right)\nonumber
\end{align}
with 
\begin{align}
	\Avgham_r = \frac{1}{2}\sqrt{4\Perturbparam^2 + \Enslabel_0^2}\op\sigma_z.
\end{align}
This differential equation can be solved analytically and yields
\begin{align}\label{eq:RotQMErho}
	\matelem{1}{\Avgdm_r(t)}{1} &= \matelem{1}{\Avgdm_r(0)}{1} \\
	\matelem{1}{\Avgdm_r(t)}{2} &= e^{-i\Escale\sqrt{4\Perturbparam^2+\Enslabel_0^2}-\Escale^2\Pwidth^2 t}\matelem{1}{\Avgdm_r(0)}{2} \label{eq:RotQMErho2}
\end{align}
We obtain immediately from Eqs.~\eqref{eq:RotQMErho},\eqref{eq:RotQMErho2} that the asymptotic state is
\begin{align}\label{eq:PureDeph}
	\lim_{t\to\infty}\matelem{1}{\Avgdm_r(t)}{1} &= \matelem{1}{\Avgdm_r(0)}{1} \;\; ;\;\;
	\lim_{t\to\infty}\matelem{1}{\Avgdm_r(t)}{2} &= 0.
\end{align}
Applying the inverse transform $R^\dagger \Avgdm_r R$, we find that the asymptotic state is the projection of the initial state onto the eigenbasis of the ensemble-averaged Hamiltonian $\Avgham$, c.f., Eq.~\eqref{eq:AsymptState}. In other words, in the eigenbasis of $\Ham_0$ the ensemble-averaged dynamics result in a dephasing process towards the basis defined by the eigenvectors of $\Avgham$. Thus, initial coherences defined with respect to the eigenbasis of $\Ham_0$ do not decay asymptotically as illustrated in Fig.~\ref{fig:QubitAsympt}.

\subsection{Effect of long-range couplings: Lattice in the strong bias limit with fully uncorrelated disorder}\label{sec:Anderson}

\begin{figure}
	\includegraphics[width = 0.5\textwidth]{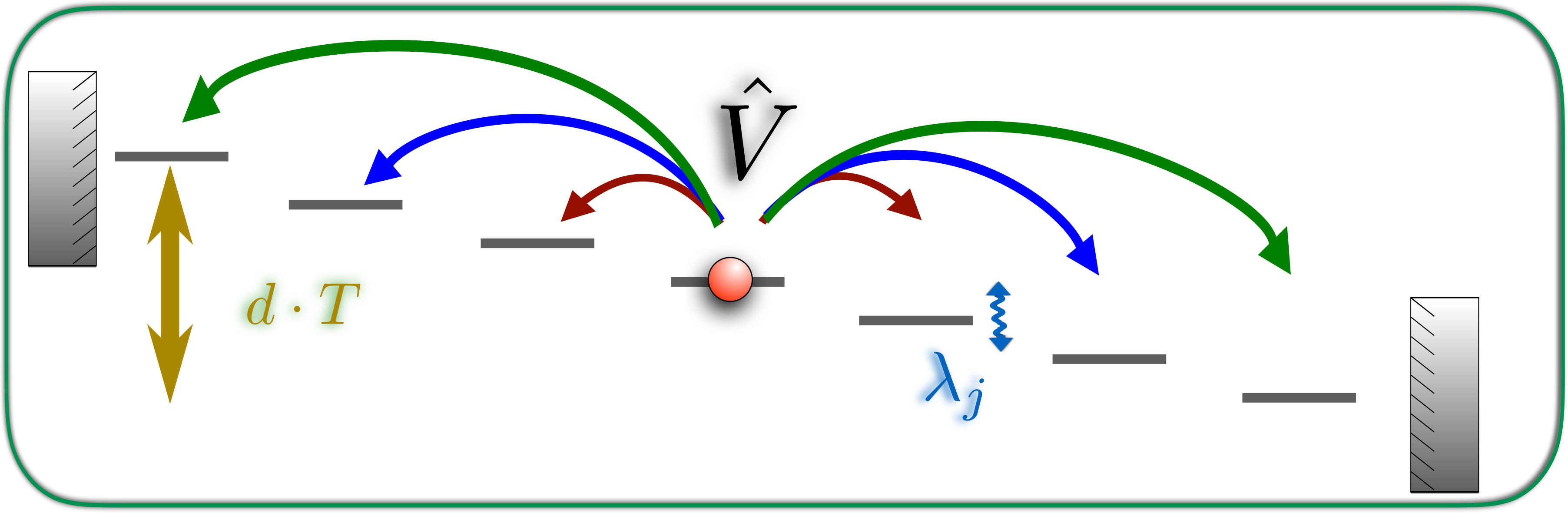}
	\caption{Illustration of a one-dimensional potential ladder with short and long-range coupling $\Perturbop$, a potential step $T$ and on-site disorder $\Enslabel_j$ (c.f. random Hamiltonians defined in Eq.~\eqref{eq:Perturb Anderson ensemble}).}
	\label{fig:TiltedAnderson}
\end{figure}

We consider a one-dimensional tight-binding model with $\Dim$ sites with on-site disorder and a constant potential ladder as illustrated in Fig.~\ref{fig:TiltedAnderson}. The random Hamiltonian ensemble is parametrized as
\begin{align}\label{eq:Perturb Anderson ensemble}
	\ens{\Ham_{\vec{\Enslabel}}= \Ham^s_{\vec{\Enslabel}}+\Perturbparam\Perturbop + \op{T} \;\; , \;\; p_{\vec{\Enslabel}}=\prod_{j=1}^\Dim p_\textrm{Ga}(\Enslabel_j)},
\end{align}
where $\Enslabel_j$ are identically independently distributed (i.i.d.) Gaussian (c.f. Eq.~\eqref{eq:GaussianDistr}) dimensionless random variables with average value $\avg\Enslabel_j=0$ and variance $\Pwidth$, and
\begin{align}\label{eq:Perturb Anderson Ham}
	\Ham_{\vec{\Enslabel}}^s &= \hbar\Escale \sum_{j=1}^\Dim \Enslabel_j \ketbra{j}{j} \;\; ; \;\; 	\op{T} = \hbar\Escale T\sum_j^{d} j\ketbra{j}{j}
\end{align}
where $\hbar\Escale$ is the reference energy, $T$ represents a constant energy shift of the levels, $\Perturbparam$ is a dimensionless perturbation parameter, and $\Perturbop$ (with $\matelem{j}{\Perturbop}{j} = 0, \forall j$) is the potential coupling. 
To study the effect of short- and long-range couplings, we consider nearest-neighbour (NN) interactions $\Perturbop_\textrm{NN} = \sum_{j=1}^{\Dim-1} \ketbra{j}{j+1}+\ketbra{j+1}{j}$ as well as dipole-type couplings with different exponents,
\begin{align}\label{eq:Couling}
	\op{V}_x = \hbar\Escale\sum_{j=1}^d\sum_{k=j+1}^d \frac{1}{\abs{k-j}^x}\ketbra{j}{k} + c.c.
\end{align} 
with $x = 0,1,3$.

This model for instance describes s-band electron transport in the presence of a strong electrical potential \cite{Kropf2019}, and corresponds to an Anderson-type model \cite{Anderson1958} in the strong bias limit. Experimentally, it can be simulated using atom-optics simulators \cite{Meier2016,Alberti2014} or photonic wave-guides \cite{Keil2012,Martin2011,Marshall2009}.  

The eigenvalues of the unperturbed part $\Ham^0_{\vec{\Enslabel}} = \Ham^s_{\vec{\Enslabel}} + \op{T} $ are $\Eno_{j,\Enslabel} =\hbar\Escale(jT + \Enslabel_j)$ (i.e. $\Eavg_j = \hbar\Escale jT$). As was derived in Section \ref{sec:PME derivation}, the corresponding first-order in $\Perturbparam$ perturbative master equation, Eq.~\eqref{eq:Perturb weak QME1}, is fully characterized by the characteristic function, Eq.~\eqref{eq:CharacFunction}, $\Charact_{jk}(\Escale t) = \textrm{exp}[-\Escale^2\Pwidth^2 t^2]$. Using the latter and Eqs.~\eqref{eq:PrateOLargeTimeS}-\eqref{eq:PrateBLargeTimeS} we obtain for the decoherence rates
\begin{align}
	\PrateO_{jk}(t) &=  -i\Escale \left(j-k\right)T -2\Escale^2\Pwidth^2 t \label{eq:Perturb Gauss PrateO}\\
	\PrateA_{jk}(t) 
	&\approx -\frac{2\Escale\Pwidth^2t}{(j-k)T} \label{eq:Perturb Gauss PrateA} \\
	\PrateB_{jkrj}(t) &\approx 0 \label{eq:Perturb Gauss PrateB}
\end{align}
where we neglected the fast decaying contribution to $\PrateA_{jk}(t)$. Here $\PrateB_{jkrj}(t) = 0$ because the eigenvalues are i.i.d. Note the factor $2$ in Eqs.~\eqref{eq:Perturb Gauss PrateO} and \eqref{eq:Perturb Gauss PrateA} as compared to the single qubit case, Eqs.~\eqref{eq:QubitPrateO} and \eqref{eq:QubitPrateA}, which is due to the i.i.d. condition.

As a numerical example, we consider a system of $\Dim=30$ sites with $\Pwidth = \Perturbparam$ and $T =10\Perturbparam$, and a broad, centered Gaussian initial state
\begin{align}\label{eq:Peturb initial gauss state}
	\Dm_0=\ketbra{G_0}{G_0}, \;\; \ket{G_0} =\frac{1}{\mathcal{N}} \sum_{j=1}^\Dim p_\textrm{Ga}'(j) \ket{j},\;\; \braket{G_0}{G_0}=1,
\end{align}
with $\mathcal{N}$ the normalization constant and $p_\textrm{Ga}'$ a Gaussian distribution with average $(\Dim + 1) / 2$ and variance $\sqrt{\Dim}$. This could for instance model a photo-excitation in a correlated thin-film transition metal-oxide heterostructure with a strong intrinsic electrical field \cite{Kropf2019}. 

\begin{figure}
	\includegraphics[width = \columnwidth]{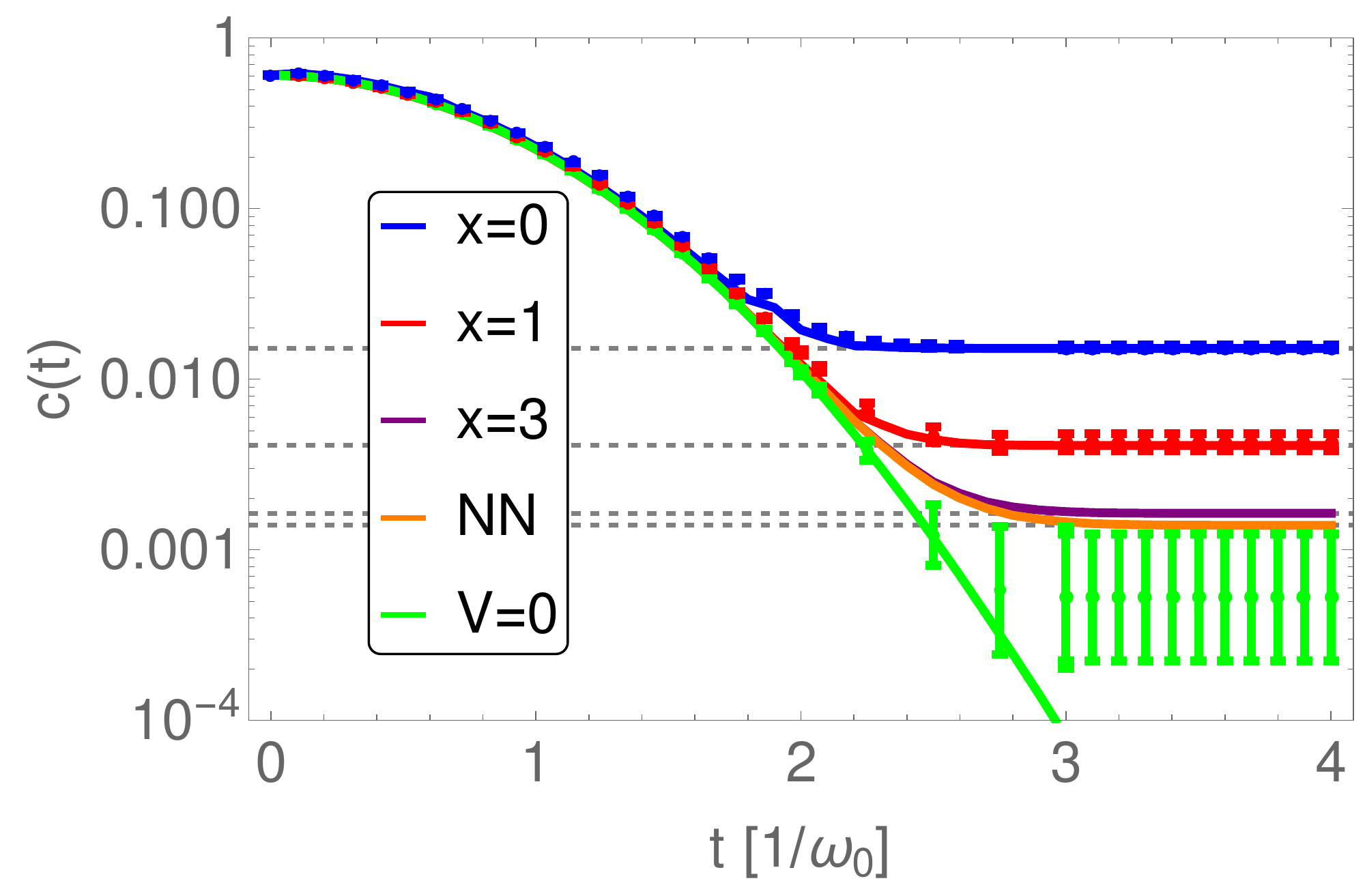}
	\caption{Evolution of the total coherence, Eq.~\eqref{eq:Coherence}, for different dipole-type coupling potentials with exponent $x=0,1,3$ (c.f. Eq.~\eqref{eq:Couling}), nearest-neighbour coupling (NN) and no coupling ($V=0$). The longer-range the coupling, the more coherences  are present in the asymptotic state characterized by Eq.~\eqref{eq:AsymptState} (horizontal gray dashed lines). Numerical integration of the perturbative master equation, Eq.~\eqref{eq:Perturb weak QME1} with rates Eq.~\eqref{eq:Perturb Gauss PrateO}-\eqref{eq:Perturb Gauss PrateB}, (solid lines) agrees with the direct numerical averaging (dots) with $10^6$ realizations of the disorder for $x=0,x=1$. For $x=3$, NN and $V=0$ (only $V=0$ is shown for visual purposes) the asymptotic state is not reached due the statistical error $ \sim \sqrt{10^6}$. The parameters are $d=30, \Pwidth = \Perturbparam, T = 10\Perturbparam$ and the initial Gaussian state, Eq.~\eqref{eq:Peturb initial gauss state}, has width $\sqrt{30}$ and average value $31/2$.
	}
	\label{fig:Anderson}
\end{figure}

In Fig.~\ref{fig:Anderson}, we show the evolution of the total coherence
\begin{align}\label{eq:Coherence}
	c(t) : = \sum_{j\neq k} \abs{\rho_{jk}}
\end{align}
for $\Perturbop_\textrm{NN}$, $\Perturbop_x$ ($x=0,1,3$), and $\Perturbop = 0$. The dynamics are obtained by numerical integration of the perturbative master equation \eqref{eq:Perturb weak QME1} with the 'NDsolve' routine in Mathematica 11.0. At short times, the coherences decay, independently of the coupling potential, as $\sim e^{-\Pwidth^2 \Escale^2 t^2}$ under the action of the rate $\PrateO_{jk}$. At large times, for any non-vanishing potential $\Perturbop$, the decay eventually stops, and a finite amount of coherence remains in the asymptotic state, which is in agreement with Eq.~\eqref{eq:AsymptState}. For long-range potentials the total coherence of the asymptotic state, $c_\infty = \lim_{ t\to \infty} c(t)$, is larger than for short-range potentials. Indeed, for a fully connected network ($x=0$) the total coherence  in the asymptotic state is $c_\infty = 15.1\times 10^{-3}$, for slowly decreasing coupling ($x = 1$) we obtain $c_\infty = 4.1\times 10^{-3}$, and for dipole-type coupling ($x = 3$) we have $c_\infty = 1.6\times 10^{-3}$. Interestingly, the nearest-neighbours coupling converges to $c_\infty = 1.4\times 10^{-3}$, which is close to the value for the dipole coupling. 

As proof of consistency of the perturbative master equation approach, we computed the full dynamics by numerical exact averaging using the numerical solver from the python Qutip package 4.3.0 \cite{Johansson2013}. For the long-range interactions $x=1$ and $x=0$, we found that $10^6$ realizations are sufficient for the convergence of the numerical averaging. The results of the master equation and direct averaging well agree as shown in Fig.~\ref{fig:Anderson}. For the shorter range interactions ($x=3$, $\Perturbop_{NN}$ and $\Perturbop = 0$), $10^6$ realizations were insufficient. In Fig.~\ref{fig:Anderson} we show the deviations for $V=0$ for which we know that the master equation is exact \cite{Kropf2016a}. Note that on the same computer the numerical integration of the master equation was ten times faster than the direct numerical averaging with $10^6$ realizations. As further test we verified that the fidelity between the density matrix from the numerical computations and the master equation is larger than $0.999$ at all times. Furthermore, we found that the relative purity $\tr{\rho^2}/\tr{\rho^2_\textrm{Num}}$ shows deviation of up to $6\%$ in the time range where the decay enters the asymptotic state ($1.5\lesssim t \lesssim 2.5$), but eventually converges to the same value (for $x = 0,1$). This deviation occurs due to the approximation in the time-dependence of the decoherence rates Eqs.~\eqref{eq:PrateOLargeTimeS}-\eqref{eq:PrateBLargeTimeS}.

Overall we find that the dephasing induced by diagonal disorder does not, in the presence of couplings $\Perturbop$ between the eigenstates, lead to a full decay of the coherences. This is in stark contrast to dynamical diagonal noise such as considered in the Hacken-Strobl model \cite{Haken1973} or for homogeneous broadening descriptions \cite{Loudon2000}. Moreover, the coherences are most efficiently protected from the disorder-induced decoherence by long-range couplings. This can be understood by analogy to the qubit case studied in Section \ref{sec:Perturb single Qubit} where we demonstrated that the ensemble-averaging leads to an effective dephasing in the eigenbasis $\{\ket{n}\}$ of the ensemble-averaged Hamiltonian $\Avgham$ (c.f. Fig.~\ref{fig:QubitAsympt}). For $\Perturbop = 0 $, the basis $\{\ket{n}\}$ is equal to the quantization basis $\{\ket{j}\}$, and we thus have a pure dephasing process so that all coherences vanish. However, the stronger the potential $\Perturbop$, the more the eigenstates of $\avg{\Ham}$ will deviate from $\ket{j}$, and thus the more the dephasing basis $\{\ket{n}\}$ differs from the quantization basis. Consequently, a larger amount of the coherences of the initial state do not decay asymptotically.

This result could be of relevance for a better understanding of the interplay between disorder and quantum coherent transport (which relies on the coherences) in strongly connected networks such as the Fenna-Matthew-Olson photo-synthesis molecular complex \cite{Walschaers2016a}, assemblies of ultra-cold Rydberg atoms \cite{Scholak2014}, strongly-correlated materials \cite{Zhang2017b, Kropf2019}, superconducting circuits \cite{Mostame2012} or photonic circuits \cite{Crespi2013}.

\subsection{Many-body (bosons) dynamics with strongly correlated noise}\label{sec:Bosons}
\begin{figure}
	\includegraphics[width = 0.23\textwidth]{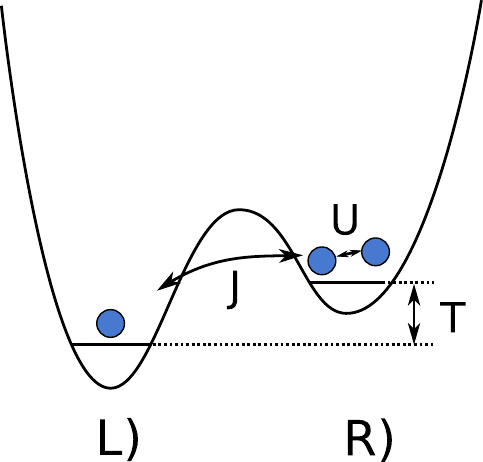}
	\caption{Double-well potential with a tilt $T$ between the wells left (L) and right (R) filled with bosons (blue dots) with on-side interaction $U$ and tunnelling barrier $J$ as an illustration of the Bose-Hubbard Hamiltonian \eqref{eq:Perturb B-H Ham}.}.
	\label{fig:Double-well}
\end{figure}

We consider an asymmetric double-well potential with $N$ interacting bosons (see illustration Fig.~\ref{fig:Double-well}) described in terms of a tilted Bose-Hubbard model \cite{Hubbard1963,Kanamori1963,Gutzwiller1963}). Such a setting can be experimentally implemented with ultracold bosons in optical lattices \cite{Jaksch1998,Morsch2006} and was studied, e.g., in the context of quantum Chaos \cite{Hiller2012}, superfluidity \cite{Gerster2016}, and distinguishability \cite{Brunner2018}. Static diagonal noise can either arise from random on-site interaction and/or from fluctuations in the tilt  between the left and right wells. In an experiment with optical lattices, the former may arise from imprecisions in the Feshbach resonances used to fix the magnitude of the interaction \cite{Inouye1998,Leggett2001}, while the former comes from fluctuations of the trapping potential. Considering one or the other is equivalent from a formal point of view, and we arbitrarily choose the interaction to be random.

Firstly, we fix the notation and write the Bose-Hubbard Hamiltonian in the absence of noise as
\begin{align}
	\Ham_{BH} = \Ham^0+\Perturbparam\Perturbop \label{eq:Perturb B-H Ham}
\end{align}
with the unperturbed kinetic term
\begin{align}
	\Ham^0 =&  \hbar \Escale T \left(\op{N}_L -\op{N}_R\right) \\
	&+ \hbar \Escale \frac{U_0}{2}\left(\op{N}_L(\op{N}_L-1)+\op{N}_R(\op{N}_R-1)\right) \nonumber,
\end{align}
and the perturbation 
\begin{align}
	\Perturbop =& -\hbar \Escale  J\left(\op{a}_L^\dagger \op{a}_R + \op{a}_R^\dagger \op{a}_L\right).
\end{align}
Here $\op{a}_L$, $\op{a}_R$, $\op{a}_L^\dagger$, $\op{a}_R^\dagger$ are the bosonic annihilation and creation operators of the left ($L$) and right ($R$) wells respectively, and the number operators $\op{N}_L = \op{a}_L^\dagger \op{a}_L$, $\op{N}_R = \op{a}_R^\dagger \op{a}_R$. The tunnelling rate is denoted as $\hbar \Escale J \in \Re^+$, the on-site interaction reads $\hbar \Escale U\in \Re$ and the tilt between the two wells is given by $\hbar \Escale T\in \Re$. The eigenvalues of the unperturbed Hamiltonian $\Ham^0$ can be written as $E_{m}^0= \hbar\Escale(\beta_m U_0 + \chi_m T)$ with the constant factors $\beta_m = 1/2m(m-1)+1/2(N-m)(N-m-1)$ and $\chi_m = (2m-N)$, $m=0,\ldots,N$ fixed by the number of bosons $N$ in the system and using the Fock basis ordering $\ket{N,0},\ket{N-1,1},\ldots\ket{0,N}$. For example for $N=3$ we have $\ket{E_1} = \ket{3,0},\ket{E_2}=\ket{2,1},\ket{E_3}=\ket{1,2},\ket{E_4} = \ket{0,3}
$ with $\beta_1=3,\beta_2=1,\beta_3=1, \beta =3$ and $\chi_1=-3,\chi_2=-1,\chi_3=1,\chi_4 =3$.

Adding a random noise $\delta U$ to the interaction, we replace $U_0 \rightarrow U_0 +\delta U$. Consequently, the  eigenvalues of the unperturbed random Hamiltonians are given by $E_{m,\Enslabel}^0= \hbar\Escale(\beta_m U_0 + \chi_m T) + \hbar \Escale\Enslabel_m$, with $\Enslabel_m = \delta U \beta_m$. The noisy eigenenergies are thus strongly correlated as they are characterized by a single random variable $\delta U$, i.e., the eigenvalues are \emph{not} independently distributed, as opposed to the previously studied model of a one-dimensional potential ladder with on-site disorder.

Assuming a Gaussian distribution $p_\textrm{Ga}(\delta U)$ of mean value $0$ and variance $\Pwidth$, the characteristic function, Eq.~\eqref{eq:CharacFunction}, evaluates to $\Charact_{jk}(\Escale t) = \textrm{exp}[-1/2\Escale^2(\beta_j-\beta_k)^2\Pwidth^2 t^2]$, and the decoherence rates Eqs.~\eqref{eq:PrateOLargeTimeS} - \eqref{eq:PrateBLargeTimeS} read
\begin{align}\label{eq:BosonPrateO}
	\PrateO_{jk}(t) = & -i\Escale\left[(\beta_j-\beta_k)U_0+(\chi_j-\chi_k)T\right] \nonumber\\
	&\quad-\Escale^2(\beta_j-\beta_k)^2\Pwidth^2 t \\ \label{eq:BosonPrateA}
	\PrateA_{jk}(t) &\approx \frac{-i\Escale(\beta_j-\beta_k)^2\Pwidth^2 t}{(\beta_j-\beta_k)U_0+(\chi_j-\chi_k)T} \\
\PrateB_{jkrj}(t) &\approx i\left(\Escale\Pwidth^2 t\frac{(\beta_r-\beta_k)^2-(\beta_j-\beta_k)^2}{(\beta_j-\beta_k)U_0+(\chi_j-\chi_k)T}\right) \label{eq:BosonPrateB}.
\end{align}
Hence, second order processes described by $\PrateB_{jkrj}(t)$ now play a role because the eigenvalues are strongly correlated. 
To guarantee that the eigenvalues $E_{m,\Enslabel}^0$ of the unperturbed random Hamiltonian remain non-degenerate for most realizations of the noise, we require $\abs{T} \gg \sigma$. For the perturbation to remain small, we further impose the condition $\alpha J < \abs{U_0 + T}$. 
\begin{figure}[t]
	\includegraphics[width = \columnwidth]{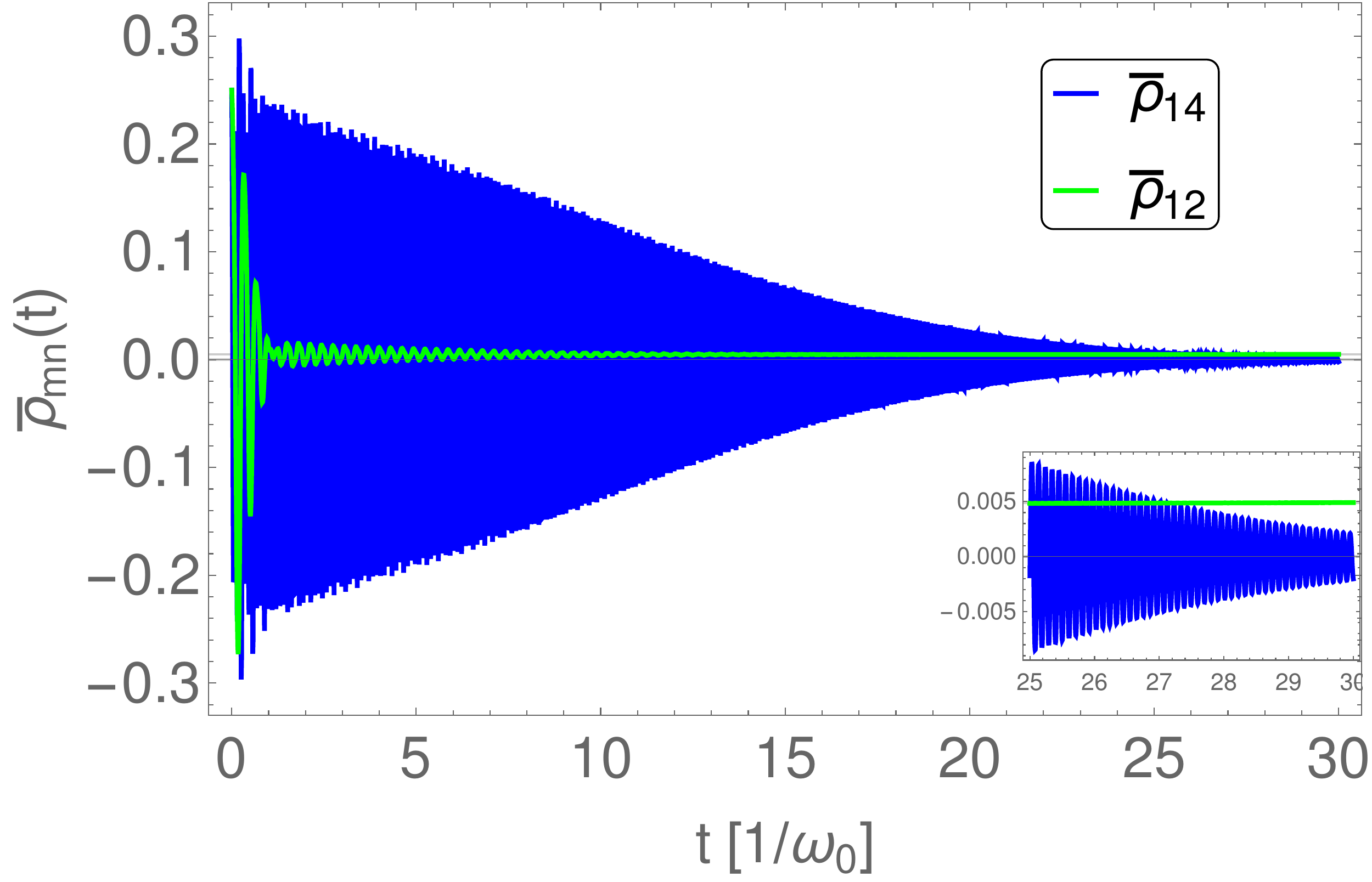}
	\caption{Evolution of ensemble-averaged coherences for the Bose-Hubbard Hamiltonian Eq.~\eqref{eq:Perturb B-H Ham} with $N=3$ bosons and random on-site interactions for the initial state $1/\sqrt{4}\sum_{j=1}^4\ket{j}$ obtained from numerical integration of the perturbative master equation. The element $\avg{\rho}_{14}(t)$ (blue line) decays to zero, but slowly because $\beta(1)-\beta(4) =0$, while $\avg{\rho}_{12}(t)$ (green line) decays fast, but to a non-vanishing value $\sim 4.87\cdot 10^{-3}$ (see inset). The parameters are $J=1$, $U_0=1$, $T=10$, $\Perturbparam =1$ and $\Pwidth =1$.
}
	\label{fig:Bosons}
\end{figure}

Interestingly, for energy levels $a,b$ that are symmetric in their number of bosons in the left/right well (e.g., $E_1 = \ket{3,0}$ and $ E_4 = \ket{0,3}$), the real parts of the first two decoherence rates, Eq.~\eqref{eq:BosonPrateA},\eqref{eq:BosonPrateB}, vanish: $\real{\PrateO_{ab}(t)}=0$ and $\real{\PrateA_{ab}(t)}=0$ (because $\beta_a-\beta_b=0$).  As a consequence, the associated off-diagonal elements $\avg{\rho}_{ab}$ evolve only from the action of the average Hamiltonian $\Avgham$ and the second-order rates $\PrateB_{abra}(t)$, which results in a slow decay. This is illustrated in Fig.\ref{fig:Bosons} for $N=3$, the initial state $1/\sqrt{4}\sum_{j=1}^4 \ket{E_j}$, and with parameters $J=1$, $U_0=1$, $T=10$, $\Perturbparam =1$ and $\Pwidth =1$. Indeed, the coherence $\avg{\rho}_{14}(t)$, for which $\beta_1-\beta_4=0$, decays ten-times slower than $\avg{\rho}_{12}(t)$, for which $\beta_1-\beta_2=-2$. However, note that since $\chi_1-\chi_4 = 0$, there is no direct coupling between the states $\ket{E_1}$ and $\ket{E_4}$ so that $\avg{\rho}_{14}(t)$ asymptotically converges to zero. On the contrary, since $V_{12} = -\sqrt{3}\hbar\Escale J$ the fast decaying coherence $\Avgdm_{12}(t)$ converges to a finite value $c_\infty \sim 4.87 \cdot 10^{-3}$ obtained from Eq.~\eqref{eq:AsymptState} (c.f. inset of Fig.~\ref{fig:Bosons}).

Hence, in addition to coherences being present in the asymptotic state due to the coupling $\Perturbop$, the symmetry of the Hamiltonian gives rise to slowly decaying coherences. This effect could be exploited to generate long-lived coherences of many-body states in systems subject to generic on-site noise. 


\section{Conclusions}\label{sec:Conclusions}

We derived general perturbative master equations to describe the ensemble-averaged dynamics of quantum systems that are subject to static noise (which includes disorder, the coupling with a classical environment, or slowly drifting experimental parameters). The pertubative master equations not only provide a description of the ensemble-averaged dynamics in terms of physically interpretable operators, energy shits, and (time-dependent) decoherence rates, but also require much less computational resources to numerically integrate as compared to direct numerical averaging of the dynamics.

We treated in details systems described by Hamiltonians with diagonal static noise, and a noise-free perturbative coupling potential. In the range of parameters where perturbation theory converges for most realizations of the noise, the first-order master equation describes the ensemble-averaged dynamics on all time scales up to systematic errors. The effect of the ensemble-averaging can be understood as a dephasing (decoherence) process in the eigenbasis of the average Hamiltonian, which was illustrated in the Bloch sphere for a two-level system with a random energy splitting. Thus, the asymptotic state is the projection of the initial state on the eigenbasis of the latter. This was numerically verified using the perturbative master equation and direct numerical averaging for several examples.

The generality of our approach was shown with a one-dimensional tight-binding model with on-site disorder and a Bose-Hubbard double-well Hamiltonian with on-site interaction noise. In the first example, we showed that the longer range the perturbative coupling potential is, the more coherences are protected from the dephasing and remain in the asymptotic state. In the second example, we showed that strong correlations in the noise distribution result in a slow decay of the coherences, which are thus partially protected from the averaging-induced decoherence. For all scenarios we discussed various experimental setups to measure these effects. Our work suggest that the later could be exploited to protect coherences for quantum applications. Conversely, static random distribution could be introduced on purpose to give rise to a desired type of dephasing, and, for instance, drive a quantum system into a target non-equilibrium quantum state.

\begin{acknowledgments}
The author thanks Gabriel Dufour, Vyacheslav Shatokhin, Andreas Buchleitner, Fausto Borgonovi, Roberto Auzzi, Dario Mazzoleni and Clemens Gneiting for interesting discussions.  C.M.K. acknowledges support by the Iniziativa Specifica INFN-DynSysMath. 
\end{acknowledgments}


\appendix

\section{Neumann series}\label{app:Neumann}

The inverse of a a given matrix $M$, which is almost equal to an invertible matrix $M_0$, in the sense that 
 \begin{align}\label{eq:NeumannCondition}
 	\limit{n}{\infty} \left(\Id - {M_0}^{-1}M\right)^n =0,
 \end{align}
 can be written as a Neumann series (see e.g. \cite{Stewart1998}, p.75)
 \begin{align}\label{eq:Perturb Neuman series}
 	M^{-1} = \sum_{n=0}^\infty \left(\Id -{M_0}^{-1}M\right)^n {M_0}^{-1}.
 \end{align}
This result can be applied to the perturbative ensemble-averaged dynamical matrix ($M=\avg{\Dmatrix}$, $M_0 = \avg{\Dmatrix^{0}}$). We remark that the zeroth order dynamical matrix $ \Dmatrix^{0}$, Eq.~\eqref{eq:Perturb F0}, is diagonal in the eigenbasis of the unperturbed Hamiltonians, and, thus, directly invertible. Moreover, $\Id - \avg{\Dmatrix^0}^{-1}\avg\Dmatrix = \avg{\Dmatrix} - \avg{\Dmatrix^{0}} \sim \alpha$ and thus condition \eqref{eq:NeumannCondition} is fulfilled and the series converges. The inverse of the ensemble-averaged dynamical matrix is then given by
\begin{align*}
	\avg{\Dmatrix}^{-1}(t) &= \sum_{m=0}^\infty \left(\Id -\avg{\Dmatrix^0}^{-1}(t)\avg\Dmatrix(t)\right)^m \avg{\Dmatrix^0}^{-1}(t) \\
	& = \sum_{m=0}^\infty \left(-\sum_{k=1}^\infty \Perturbparam^k \avg{\Dmatrix^0}^{-1}(t) \avg{\Dmatrix^k}(t) \right)^m\avg{\Dmatrix^0}^{-1}(t),
\end{align*}
and the master equation matrix, Eq.~\eqref{eq:QmatrixDef}, then reads 
\begin{align*}
	\avg{\QMEmatrix}(t) =& \dot{\avg{\Dmatrix}}(t)\avg{\Dmatrix}^{-1}(t)\\
	=&\left(\sum_{n=0}^\infty \Perturbparam^n \dot{\avg{\Dmatrix^n}}(t)\right) \\ &\cdot \sum_{m=0}^\infty \left(-\sum_{k=1}^\infty \Perturbparam^k \avg{\Dmatrix^0}^{-1}(t) \avg{\Dmatrix^k}(t) \right)^m\avg{\Dmatrix^0}^{-1}(t).
\end{align*}

\section{First-order Perturbation equation}\label{sec:Perturb weak PME derivation}

Here we show the computations for deriving the master equation \eqref{eq:Perturb weak QME1}. For more details see \cite{Kropf2017a}.
 \paragraph{Zeroth order $\Perturbparam^0$:}
 The zeroth order dynamical matrix captures the dynamics of solely the unperturbed part, and describes the evolution of the coherences in the eigenbasis $\ens{\ket{j}}$. We recall that $\op{U}_\Enslabel^0(t) = e^{-\frac{it}{\hbar}\Ham_\Enslabel^0} = \sum_j e^{-\frac{it}{\hbar}E_{\Enslabel,j}} \ketbra{j}{j}$ and derive the ensemble-average value
\begin{align}\label{eq:Perturb F0}
\dot{\avg{F^0}}(t)&=  \sum_{jk,rs}^\Dim \Lioumatelem{jk}{\avg{\mathcal{U}^0_\Enslabel} (t)}{rs} = 
\sum_{j,k=1}^\Dim\dot{\avg{\Phase}}_{jk}(t) \Liouketbra{jk}{jk} 
\end{align}
 where the ensemble-averaged phase factors
\begin{align}\label{eq:Perturb avg phases}
 	\avg{\Phase}_{jk}(t) \equiv \int d{\vec{\Enslabel}} p_{\vec{\Enslabel}} e^{-\frac{it}{\hbar}(E_{\Enslabel,j}-E_{\Enslabel,k})} = \int d{\vec{\Enslabel}} p_{\vec{\Enslabel}} e^{-it\Escale(\Enslabel_j-\Enslabel_k)},
 \end{align}
 and thus
 \begin{align}
 \avg{F^0}^{-1}(t) =  \sum_{j,k=1}^\Dim\frac{1}{\avg{\Phase}_{jk}(t)} \Liouketbra{jk}{jk}. \label{eq:Perturb F0 inverse}
\end{align}
Note that one must remain careful here because even tough $\Phase_{jk}^{-1}(t) = \Phase_{jk}^*(t)$, in general $\avg{\Phase}_{jk}^{-1}(t) \neq \avg{\Phase}_{jk}^*(t)$. 

  \paragraph{First order $\Perturbparam^1$:}
From Eqs.~\eqref{eq:Perturb F Liouville} and \eqref{eq:Perturb average F expansion} we derive 
 \begin{align*}
 		i\hbar\avg{\Dmatrix^1}(t)&= -\frac{i}{\hbar}\int_0^t \, dt' \, \Liouketbra{jk}{jk}\avg{\mathcal{U}^0(t-t') \mathcal{\Perturb} \mathcal{U}^0(t')}\Liouketbra{rs}{rs} \nonumber \\
 	& =\sum_{\mathclap{\substack{j,k=1\\ j\neq k}}}^\Dim  f_{jk}(t)  \Perturb_{kj}  \left(\Liouketbra{kk}{jk}- \Liouketbra{jj}{jk} \right) \\&\quad\quad + f_{jk}(t)  \Perturb_{jk}  \left(\Liouketbra{jk}{kk}- \Liouketbra{jk}{jj} \right)    \numberthis \label{eq:Perturb F1 with f} \\
 	&+\sum_{\mathclap{\substack{j,k,r=1\\ j\neq r\neq k}}}^\Dim f_{jkrj}(t) \Perturb_{jr} \Liouketbra{jk}{rk} -f_{jkkr}(t)  \Perturb_{rk} \Liouketbra{jk}{jr},
\end{align*}
where we defined for simplicity of reading
\begin{align*}
	f_{jk}(t) \eqdef \int_0^t dt' \, \avg{\Phase}_{jk}(t') \;\; , \;\; f_{jkrj}(t) \eqdef \avg{\Phase_{jk}(t)\int_0^t dt' \,\Phase_{rj}(t')},
\end{align*} 
 and where for $f_{jkrj}$ the ordering of the indices refers to the ordering of the indices of the phase factors.
 Note that in the above above, we separated out the terms for which the phase factors are equal to the identity prior to the ensemble-averaging because in general $\avg{\Phase_{jk}(t)}\delta_{j,k} \neq \avg{\Phase_{jj}(t)}=1 $. 

Using the zeroth Eqs.~\eqref{eq:Perturb F0} and \eqref{eq:Perturb F0 inverse} and the first order Eq.~\eqref{eq:Perturb F1 with f} for the dynamical matrix, we can compute the first-order expansion of $\avg{\QMEmatrix}$, Eq.~\eqref{eq:QmatrixDef}, which fully characterizes the perturbative master equation. To zeroth order we obtain
\begin{align}
	\avg{\QMEmatrix^0}&=\dot{\avg{\Dmatrix^0}} \avg{\Dmatrix^0}^{-1} 
	=  \sum_{j,k=1}^\Dim\frac{\dot{\avg{\Phase}}_{jk}(t)}{\avg{\Phase}_{jk}(t)} \Liouketbra{jk}{jk}\nonumber\\ &\eqdef  \sum_\subsum{j,k=1}{j\neq k}^\Dim\PrateO_{jk}(t)\Liouketbra{jk}{jk},\label{eq:Perturb weak Q0}
\end{align}
where we defined 
\begin{align}
 \PrateO_{jk}(t) \eqdef \frac{d}{dt} \ln \left[\avg{\Phase}_{jk}(t)\right] .
\end{align}
The first order yields
\begin{align*}
	&\avg\QMEmatrix^1(t) \equiv\dot{\avg{\Dmatrix^1}}(t)\avg{\Dmatrix^0}^{-1}(t) -\dot{\avg{\Dmatrix^0}}(t)\avg{\Dmatrix^0}^{-1}(t)\avg{\Dmatrix^1}(t)\avg{\Dmatrix^0}^{-1}(t) \\
	=&-\frac{i}{\hbar}\sum_{\mathclap{\substack{j,k=1\\ j\neq k}}}^\Dim  \Perturb_{kj} \left(\Liouketbra{kk}{jk}- \Liouketbra{jj}{jk} \right) \\
	&+\frac{1}{\hbar}\sum_{\mathclap{\substack{j,k=1\\ j\neq k}}}^\Dim \tilde\PrateA_{jk}(t)\Perturb_{jk}  \left(\Liouketbra{jk}{kk}- \Liouketbra{jk}{jj} \right)  \numberthis \label{eq:Perturb weak Q1} \\
 	&+\frac{1}{\hbar}\sum_{\mathclap{\substack{j,k,r=1\\ j\neq r\neq k}}}^\Dim \tilde\PrateB_{jkrj}(t)\Perturb_{jr} \Liouketbra{jk}{rk} -\tilde\PrateB_{jkrj}^*(t) \Perturb_{rj} \Liouketbra{kj}{kr}. 
\end{align*}
where we defined the anti-Hermitian rate matrix
\begin{align}
	\tilde\PrateA_{jk}(t) \eqdef& -i \left(\dot{f}_{jk}(t) -\PrateO_{jk}(t)f_{jk}(t)\right)\nonumber \\
	=& -i\left[\avg{\Phase}_{jk}(t) - \PrateO_{jk}(t)\int_0^t dt' \avg{\Phase}_{jk}(t')\right] = -\tilde\PrateA_{kj}^*(t),
\end{align}
and the matrix
\begin{align*}
	\tilde\PrateB_{jkrj}(t)& \eqdef -i \left[\frac{\dot{f}_{jkrj}(t)}{\avg{\Phase}_{rk}(t)} -\PrateO_{jk}(t)\frac{f_{jkrj}(t)}{\avg{\Phase}_{rk}(t)} \right]\numberthis  \\
	=&-i - \frac{i}{\avg{\Phase}_{rk}(t)} \left( \avg{\dot{\Phase}_{jk}(t)\int_0^t dt' \,\Phase_{rj}(t')}\right) \\
	& +i \frac{\dot{\avg{\Phase}}_{jk}(t)}{\avg{\Phase}_{jk}(t)\avg{\Phase}_{rk}(t)} \avg{\Phase_{jk}(t)\int_0^t dt' \,\Phase_{rj}(t')} = -\tilde\PrateB_{kjjr}^*(t).
\end{align*}
The symmetry $\tilde\PrateB_{jkrj}=-\tilde\PrateB_{kjjr}^*$ is obtained by permuting the first two indices, $jk\rightarrow kj$, and the last two indices, $rj\rightarrow jr$, and reflects the Hermiticity of the density matrix.

In order to derive a master equation in Lindblad form following the general method described in \cite{Kropf2016a}, we should now expand $\avg\QMEmatrix^1(t)$ in a basis of Hermitian traceless operators and collect the terms for the coherent and incoherent parts. However, this proves to be, in general, technically rather involved. We prefer to go back to the Hilbert space representation using the identity
\begin{align*}
	\dot{\Avgdm}(t) = \sum_{j,k,r,s=1}^\Dim \avg\QMEmatrix_{jk,rs}(t) \ketbra{j}{r}\Avgdm\ketbra{s}{k},
\end{align*}
 in order to obtain the evolution equation in terms of the matrix elements of the ensemble-averaged density matrix. A nice form of the equation is obtained by defining
\begin{align}\label{eq:Perturb Prate AB tilde}
	\PrateA_{jk}(t) \eqdef \tilde\PrateA_{jk}(t) + i  \;\; , \;\; \PrateB_{jkrj}(t) \eqdef \tilde\PrateB_{jkrj}(t) + i, 
\end{align}
thereby isolating the time-independent parts $\PrateA(0) = -i$ and $\PrateB(0) = -i$, and with both $\PrateA(t)$ and $\PrateB(t)$ being at least of second order in time. Then, the master equation can be expressed as
\begin{align*}
			&\dot{\Avgdm}(t) = \sum_{j,k=1}^\Dim \PrateO_{jk}(t) \Projectionop_{jj} \Avgdm(t) \Projectionop_{kk}-\Perturbparam\frac{i}{\hbar} \commutator{\Perturbop}{\Avgdm(t)} \\
	& +\frac{\Perturbparam}{\hbar} \sum_\subsum{j,k=1}{j\neq k}^\Dim \PrateA_{jk}(t) \Perturb_{jk} \left[\Projectionop_{jk}\Avgdm(t)\Projectionop_{kk}-\Projectionop_{jj}\Avgdm(t)\Projectionop_{jk}\right] \numberthis \\
		& +\frac{\Perturbparam}{\hbar} \sum_\subsum{j,k,r=1}{j\neq k\neq r}^\Dim \PrateB_{jkrj}(t) \Perturb_{jr} \Projectionop_{jr}\Avgdm(t)\Projectionop_{kk} + \PrateB_{jkrj}^*(t)\Perturb_{rj} \Projectionop_{kk}\Avgdm(t)\Projectionop_{rj} \\
		&+O(\Perturbparam^2), 
\end{align*}
with the diadic operators $\Projectionop_{jk} := \ketbra{j}{k}$.

\section{Symmetric distributions}\label{app:Rates}
Let us derive the decoherence rates $\PrateA_{jk}(t)$,Eq.~\eqref{eq:Perturb PrateA def}, and $\PrateB_{jkrj}(t)$, Eq.~\eqref{eq:Perturb PrateB def}, in the case that the joint-noise-distribution $q_{jk}(\Enslabel_j-\Enslabel_k)$ for any pair of variables $\Enslabel_j,\Enslabel_k$ has vanishing odd cumulants (except the central value), or in other words, the probability density functions $q_{jk}$ are symmetric around their central value. In this case, considering furthermore that the probability distributions shall be continuous and smooth, we know that the associated characteristic function vanishes in the limit of large times,
\begin{align}
	\phi_{jk}(t) = \int d \Enslabel_j \int \Enslabel_k q_{jk}(\Enslabel_j-\Enslabel_k)e^{-i (\Enslabel_j-\Enslabel_k) t} \overset{t\rightarrow \infty}{\longrightarrow} 0 .
\end{align}

We begin with the rate 
\begin{align}\label{eq:PrateAapp}
\PrateA_{jk}(t) =i-i\avg{\Phase}_{jk}(t) + i\PrateO_{jk}(t)\int_0^t dt' \avg{\Phase}_{jk}(t'),
\end{align}
and first evaluate the integral ~\\[-13pt]
\begin{align*}
	&\int_0^t dt' \avg{\Phase}_{jk}(t')  = \int_0^t dt' e^{-\frac{i t'}{\hbar}(\Eavg_j-\Eavg_k)} \Charact_{jk}(\Escale t') \\
	&=  \int d\Delta_{jk} q_{jk}(\Delta_{jk}) \int_0^t dt' e^{-i\Escale t'(\Enslabel_j-\Enslabel_k)}e^{-\frac{i t'}{\hbar}(\Eavg_j-\Eavg_k)}\\
	&= \int d\Delta_{jk} q_{jk}(\Delta_{jk}) i\frac{e^{-i\Escale(\Enslabel_j-\Enslabel_k) t}-1}{\Escale(\Enslabel_j-\Enslabel_k) + (\Eavg_j/\hbar-\Eavg_k/\hbar)}. 
\end{align*}

Since we are working in the non-degenerate perturbation regime, i.e., $\Eno_{\Enslabel,j} - \Eno_{\Enslabel,k} = (\Eavg_j/\hbar +\Escale\Enslabel_j) - (\Eavg_k/\hbar - \Escale \Enslabel_k)\gg \Perturbparam$ for most $\Enslabel$, the probability of degenerate levels must be sufficiently small.  Thus, we can approximate the integral as
\begin{align*}
		&\int_0^t dt' \avg{\Phase}_{jk}(t') \approx \int d\Delta_{jk} q_{jk}(\Delta_{jk}) i\frac{e^{-i\Escale(\Enslabel_j-\Enslabel_k) t}-1}{\Escale(\avg\Enslabel_j-\avg\Enslabel_k) + (\Eavg_j/\hbar-\Eavg_k/\hbar)} \\
	   &= \frac{i}{\avg{\Eno_j}/\hbar-\avg{\Eno_k}/\hbar}(\avg{\Phase}_{jk}(t)-1) \\
	   & =\frac{\avg{\Phase}_{jk}(t)-1}{i\imag{\PrateO_{jk}(t)}} \label{eq:TimeIntegralPhase}
\end{align*}
Inserting this results in to the definition Eq.~\eqref{eq:PrateAapp} we obtain
\begin{align*}
	\PrateA_{jk}(t) &\approx i-i\avg{\Phase}_{jk}(t) +i \PrateO_{jk}(t) \frac{(\avg{\Phase}_{jk}(t)-1)  }{i\imag{\PrateO_{jk}(t)}}\\
	&= (\avg{\Phase}_{jk}(t)-1) \frac{ \real{\PrateO_{jk}(t)}}{\imag{\PrateO_{jk}(t)}}.
\end{align*}

For the second rate,
\begin{align*}
	\PrateB_{jkrj}(t) &= i\Biggl[\frac{\dot{\avg{\Phase}}_{jk}(t)}{\avg{\Phase}_{jk}(t)\avg{\Phase}_{rk}(t)} \avg{\Phase_{jk}(t)\int_0^t dt' \,\Phase_{rj}(t')} \\
	&\quad- \frac{1}{\avg{\Phase}_{rk}(t)} \left( \avg{\dot{\Phase}_{jk}(t)\int_0^t dt' \,\Phase_{rj}(t')}\right)\Biggr],
\end{align*}
we proceed analogously. First, we obtain 
\begin{align*}
	&\avg{\Phase_{jk}(t)\int_0^t dt' \,\Phase_{rj}(t')} \approx i\frac{\avg{\Phase}_{rk}(t) -\avg{\Phase}_{jk}}{\avg{\Eno_j}/\hbar-\avg{\Eno_k}/\hbar},
\end{align*}
where we again made use of the fact that levels coupled by the perturbation $\op{V}$ must have a vanishing probability to be degenerate. Furthermore,
\begin{align*}
& \avg{\dot{\Phase}_{jk}(t)\int_0^t dt' \,\Phase_{rj}(t')}\\
& =-\avg{\Phase}_{rk}(t)+  \frac{d}{dt}\left( \avg{\Phase_{jk}(t)\int_0^t dt' \,\Phase_{rj}(t')}\right) \\
&= -\avg{\Phase}_{rk}(t)+ i\frac{\dot{\avg{\Phase}}_{rk}(t) -\dot{\avg{\Phase}}_{jk}}{\avg{\Eno_j}/\hbar-\avg{\Eno_k}/\hbar}
\end{align*}
 We then obtain
\begin{align*}
	\PrateB_{jkrj}(t) \approx  \frac{\PrateO_{jk}(t)-\PrateO_{rk}(t)}{\avg{\Eno_j}/\hbar-\avg{\Eno_k}/\hbar}+i 
	= \frac{\real{\PrateO_{jk}(t)}-\real{\PrateO_{rk}(t)}}{\avg{\Eno_j}/\hbar-\avg{\Eno_k}/\hbar}.
\end{align*}


\bibliographystyle{apsrev4-1}
\bibliography{bibliography_PerturbDisorder}

\end{document}